\documentclass{emulateapj}

\newcommand{\vtan}{$V_{tan}$}
\newcommand{\kms}{km~s$^{-1}$}

\newcommand{\ldl}{$\lambda/{\Delta}{\lambda}$}
\newcommand{\lbol}{$\log_{10}{L_{bol}/L_{\sun}}$}

\newcommand{\ki}{\ion{K}{1}}
\newcommand{\nai}{\ion{Na}{1}}

\newcommand{\meth}{CH$_4$}

\newcommand{\wat}{H$_2$O}
\newcommand{\name}{2MASS~J03202839$-$0446358}
\newcommand{\namesh}{2MASS~J0320$-$0446}

\slugcomment{Submitted to ApJ 4 January 2008; Accepted 3 March 2008}

\shorttitle{The Unresolved Binary 2MASS~J0320$-$0446}
\shortauthors{Burgasser et al.}

\begin{document}

\title{Subtle Signatures of Multiplicity in Late-type Dwarf Spectra: The Unresolved
M8.5 + T5 Binary 2MASS~J03202839$-$0446358\altaffilmark{1}}

\author{Adam J.\ Burgasser,\altaffilmark{2,3}
Michael C.\ Liu,\altaffilmark{4,5}
Michael J.\ Ireland,\altaffilmark{6}
Kelle L.\ Cruz,\altaffilmark{3,7,8}
and Trent J.\ Dupuy\altaffilmark{4}}

\altaffiltext{1}{Some of the data presented herein were obtained at the
  W.M. Keck Observatory, which is operated as a scientific partnership
  among the California Institute of Technology, the University of
  California, and the National Aeronautics and Space Administration. The
  Observatory was made possible by the generous financial support of the
  W.M. Keck Foundation.}
\altaffiltext{2}{Massachusetts Institute of Technology, Kavli Institute for Astrophysics and Space Research, Building 37, Room 664B, 77 Massachusetts Avenue, Cambridge, MA 02139, USA; ajb@mit.edu}
\altaffiltext{3}{Visiting Astronomer at the Infrared Telescope Facility, which is operated by
the University of Hawai`i under Cooperative Agreement NCC 5-538 with the National Aeronautics
and Space Administration, Office of Space Science, Planetary Astronomy Program}
\altaffiltext{4}{Institute for Astronomy, University of Hawai`i, 2680 Woodlawn Drive, Honolulu, HI 96822, USA}
\altaffiltext{5}{Alfred P. Sloan Research Fellow}
\altaffiltext{6}{Division of Geological and Planetary Sciences, California Institute of Technology, Pasadena, CA 91106, USA}
\altaffiltext{7}{Division of Astronomy and Astrophysics, California Institute of Technology, Pasadena, CA 91106, USA}
\altaffiltext{8}{Spitzer Space Telescope Postdoctoral Fellow}

\begin{abstract}
Evidence is presented that 2MASS~J03202839$-$0446358, 
a late-type dwarf with 
discrepant optical (M8:) and near-infrared (L1) spectral types, 
is an as-yet unresolved stellar/brown dwarf 
binary with late-type M dwarf and T dwarf components.  
This conclusion is based on low-resolution, near-infrared spectroscopy
that reveals a subtle but distinctive absorption
feature at 1.6~$\micron$.  The feature, which is  also present
in the combined light spectrum of the M8.5 + T6 binary SCR 1845$-$6357, arises
from the combination of FeH absorption from an M8.5 primary and pseudo-continuum
flux from a T5$\pm$1 secondary, as ascertained from 
binary spectral templates constructed from empirical data.
The binary templates provide a far superior match to the overall near-infrared
spectral energy distribution of 2MASS~J0320$-$0446 than any
single comparison spectra.
Laser guide star adaptive optics (LGS AO)
imaging observations, including the first application of LGS AO 
aperture mask interferometry, fail to resolve a faint companion,
restricting the projected separation of the system to less than 8.3~AU
at the time of observation.
2MASS~J0320$-$0446 is the second very low mass
binary to be identified from unresolved, low-resolution, 
near-infrared spectroscopy, a technique that complements 
traditional high resolution imaging and spectroscopic methods.
\end{abstract}


\section{Introduction}

The optical and near-infrared spectral energy distributions of 
very low mass stars and brown dwarfs---late-type M, L and T dwarfs---are 
distinctly non-blackbody.  
Overlapping molecular bands and
strong line emission produce a rich array
of spectral diagnostics for classification
and characterization of physical properties.  
Considerable effort is now being
devoted toward decrypting the spectral fingerprints of late-type
dwarfs to determine masses, ages, metallicities and other 
fundamental parameters
(e.g., \citealt{luh99,gor03,metgrav,all07,liu07}).
In some cases, spectral peculiarities arise when an observed
source is in fact an unresolved multiple system, with components
of different masses, effective temperatures,
and other spectral properties.  While several classes of
stellar multiples are recognized on the basis
of their unusual spectral or photometric properties
(U Geminorum stars, M dwarf + white dwarf systems, etc.), 
identifying such cases amongst late-type
dwarfs is complicated by the influence of other physical
effects.
Delineation of spectral peculiarities that arise purely from multiplicity as opposed
to other physical effects 
is essential if we hope to unambiguously
characterize the physical properties of the lowest luminosity stars
and brown dwarfs.

Very low mass multiple systems are important 
in their own right, as they enable mass and occasionally radius
measurements (e.g., \citealt{lan01,zap04,sta06,liu08}), provide constraints for
star/brown dwarf formation scenarios (e.g., \citealt{clo03,pall07,luh07ppv})
and facilitate detailed studies of atmospheric properties
(e.g., \citealt{mehst2,liu06,mar06}).  
Of the roughly 90
very low mass multiple systems currently known,\footnote{A current list
is maintained by N.\ Siegler at \url{http://www.vlmbinaries.org}.} the majority
have been identified through high angular resolution imaging, 
using the {\em Hubble Space Telescope}
({\em HST}; e.g., \citealt{mar99a,bou03,mehst2,rei06}), 
ground-based adaptive optics systems 
(e.g., \citealt{clo03,cha04,sie03,sie05,kra05,liu06,loo08})
and more recently aperture masking interferometry (e.g., \citealt{ire08,kra08}).
However, as the vast majority of very low mass binaries have
small separations ($>$90\% have $\rho$ $<$ 20~AU; \citealt{me06ppv}),
expanding the population of known binaries to greater distances
requires either finer angular sampling or the identification of systems that
are unresolved.  The frequency of nearby, tightly-bound binaries is
also essential for a
complete assessment of the overall very low mass dwarf binary fraction, since
imaging studies provide only a lower limit to this fundamental statistic.  
Such systems are also more likely to 
eclipse, enabling radius measurements and fundamental tests of evolutionary
models (e.g., \citealt{sta06}).
While searches for radial velocity variability via high
resolution spectroscopy can be useful in 
this regime (e.g., \citealt{bas99,ken03,bas06,bla07,joe07}), 
in many cases very low luminosity and/or distant late-type dwarfs
are simply too faint to be followed up in this manner.

Recently, \citet{me0805} demonstrated that
in certain cases
the presence of an unresolved companion can be inferred
directly from the
morphology of a source's low-resolution near-infrared spectrum.
In particular, it was shown that the spectrum of the peculiar L dwarf 
SDSS~J080531.84+481233.0 (hereafter SDSS~J0805+4812; \citealt{haw02,kna04}),
which has highly discrepant optical and near-infrared spectral
classifications,
could be accurately reproduced as a combination of ``normal''
L4.5 + T5 components.
Indeed, the binary hypothesis provides a far simpler and 
more consistent explanation
for the unusual optical, near-infrared and mid-infrared properties 
of SDSS~J0805+4812 than other alternatives 
(e.g., \citealt{kna04,fol07,leg07}).  
The identification of unresolved multiples like SDSS~J0805+4812 
by low-resolution near-infrared
spectroscopy is a potential boon for low-mass multiplicity studies, as 
this method is not subject to the same physical or projected separation 
limitations inherent to high-resolution imaging and spectroscopic
techniques.  

This article reports the discovery of a second
unresolved very low mass binary system, 
{\name} (hereafter {\namesh}), identified by the morphology of its
low-resolution, near-infrared spectrum. The spectroscopic observations
leading to this conclusion are described in
$\S$~2, as are laser guide star adaptive optics (LGS AO) imaging 
observations aimed at searching for a faint companion.
Analysis of the spectral data
using the binary template matching technique
described in \citet{me0805} is presented in $\S$~3.
$\S$~4 discusses the viability of {\namesh} being a binary, with 
specific comparison to the
known M dwarf + T dwarf system SCR 1845$-$6357.  We also constrain
the projected separation
of the {\namesh} system based on our imaging observations, and discuss overall
limitations on the variety of unresolved M dwarf + T dwarf
binaries that can be identified from composite
near-infrared spectroscopy.
Conclusions are summarized in $\S$~5.

\section{Observations}

\subsection{Previous Observations of {\namesh}}

\begin{figure}
\epsscale{1.0}
\plotone{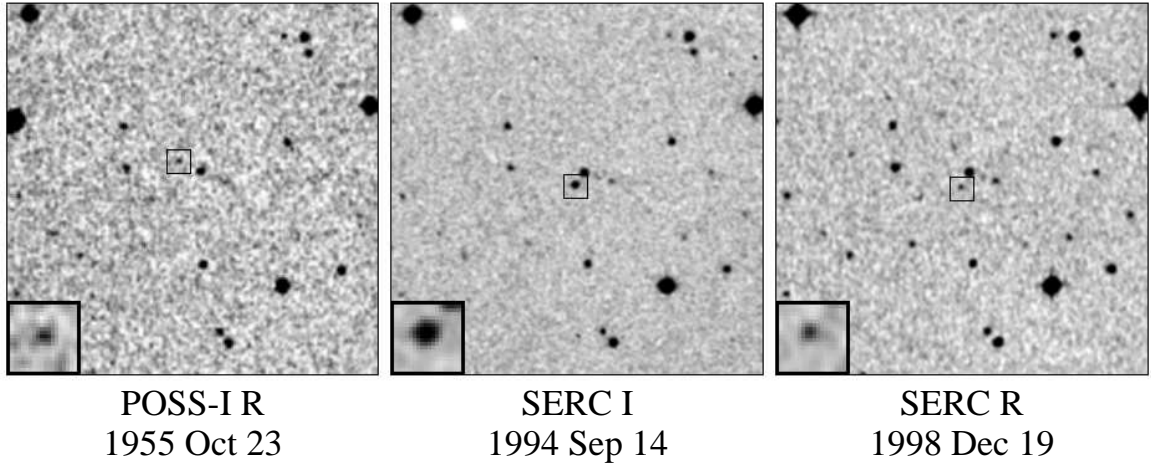}
\caption{Field images of {\namesh} from ESO $R$ (left), 
SERC $I_N$ (middle) and SERC $R$ (right) photographic plates.
All images are scaled to the same spatial
resolution, are 5$\arcmin$ on a side, and are 
oriented with north up and east to the left.  
Inset boxes 20$\arcsec$$\times$20$\arcsec$ in size indicate the 
position of the source after correcting for its motion 
($\mu = 0{\farcs}562{\pm}0{\farcs}005$~yr$^{-1}$ at position angle
$\theta = 205{\fdg}9{\pm}0{\fdg}5$) and are expanded
in the lower left corner of each image.
\label{fig_chart}}
\end{figure}

{\namesh} was originally discovered by 
\citet{cru03} and \citet{wil03} in the Two Micron All Sky Survey
(2MASS; \citealt{skr06}), and classified
M8: (uncertain) and L0.5 on the basis of optical and near-infrared
spectroscopy, respectively. The M8: optical classification is uncertain
because of the low signal-to-noise of the spectral data, and is not due to any specific
spectral peculiarity.
\citet{cru03} estimate a distance of 26$\pm$4~pc for this source
based on its classification and empirical $M_J$/spectral type
relations. \citet{dea05}, using $I$-band plate data
from the SuperCosmos Sky Survey (SSS; \citealt{ham01a,ham01b,ham01c}),
report a relatively high proper motion of 0$\farcs$68$\pm$0$\farcs$04~yr$^{-1}$ at
position angle 191$\degr$ for this source.  Figure~\ref{fig_chart} shows the field around
{\namesh} imaged by $R$ and $I$ photographic plates. 
A faint source is seen in the 1955 Palomar Sky Survey I \citep{abe59}
$R$-band image roughly at the offset position
indicated by the \citet{dea05} proper motion.  By including this source
position along with additional astrometry drawn from the SSS 
and 2MASS catalogs, an 
improved proper motion measurement of 
0$\farcs$562$\pm$0$\farcs$005~yr$^{-1}$ at
position angle 205.9$\pm$0.5$\degr$ was determined.  This proper motion
and the estimated distance indicates a rather large
tangential space velocity for {\namesh}, {\vtan} = 69$\pm$11~{\kms}, suggesting that 
it could be an older disk star.  None of the previous
studies of {\namesh} report the presence of a faint companion.

\subsection{Near-Infrared Spectroscopy}

Low resolution near-infrared
spectral data for {\namesh} were
obtained on 2007 September 16 (UT) using the SpeX spectrograph \citep{ray03}
mounted on the 3m NASA Infrared Telescope Facility (IRTF).
The conditions on this night were poor with patchy clouds,
cirrus and average seeing (0$\farcs$8 at $J$-band), and 
{\namesh} was observed as a bright
backup target ($J$ = 12.13$\pm$0.03).  
The 0$\farcs$5 slit was used to obtain 0.7--2.5~$\micron$
spectroscopy with resolution {\ldl} $\approx 120$
and dispersion across the chip of 20--30~{\AA}~pixel$^{-1}$.
To mitigate the effects of differential refraction, the slit was aligned
to the parallactic angle. Six exposures of 
90~s each were obtained 
in an ABBA dither pattern along the slit.
The A0~V star HD~18571 was observed immediately
after {\namesh} and at a similar airmass (1.21) for flux calibration.
Internal flat field and argon arc lamps were observed after
both target and flux standard observations
for pixel response and wavelength calibration.
Data were reduced with the IDL SpeXtool package, version 3.4
\citep{cus04,vac03}, using standard settings.  A detailed description of
the reduction
procedures is given in \citet{me0805}.

\begin{figure}
\epsscale{0.8}
\plotone{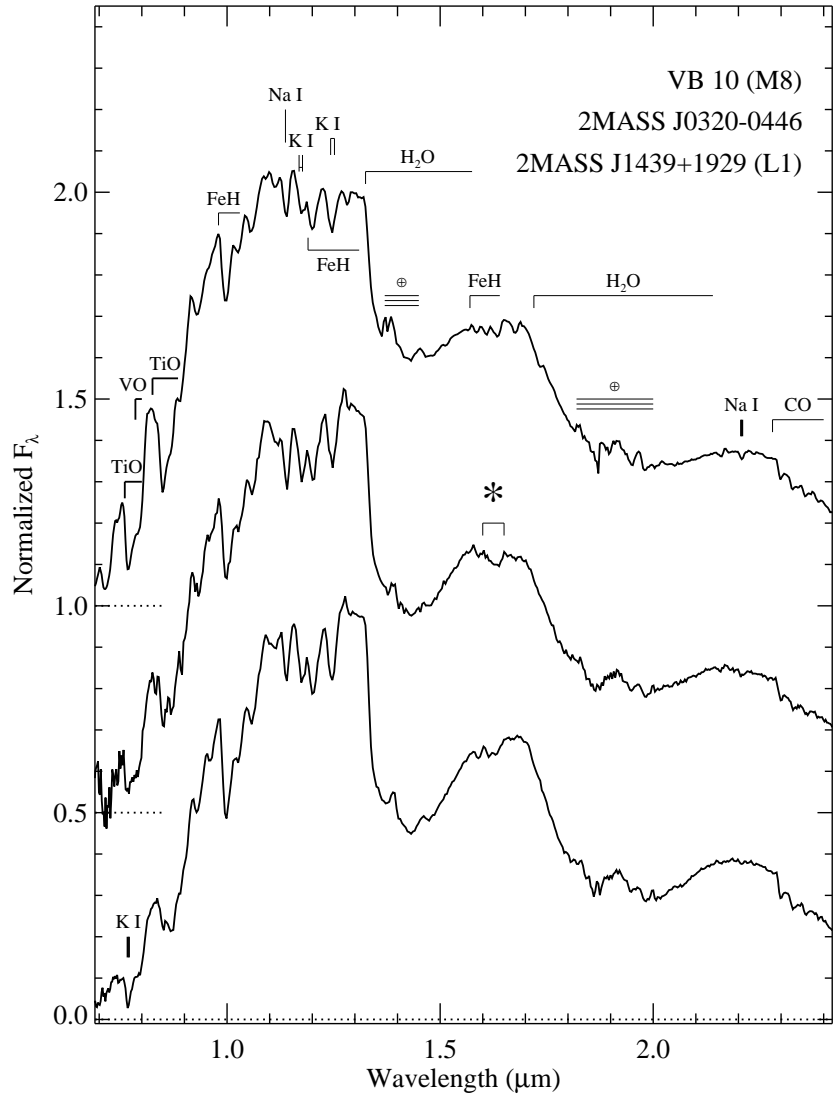}
\caption{SpeX prism spectrum for {\namesh} (center) compared
to equivalent data for VB~10 (M8)
and 2MASS~J1439+1929 (L1; see Table~\ref{tab_templates}). Spectra
are normalized at 1.25~$\micron$ and offset by
constants (dotted lines).  Prominent features resolved by these
spectra are indicated. The peculiar 1.6~$\micron$ feature in the spectrum
of {\namesh} discussed in the text is indicated by an asterisk.
\label{fig_nirspec}}
\end{figure}

The near-infrared spectrum of {\namesh} is shown in Figure~\ref{fig_nirspec}, 
compared to equivalent data for the optical spectral standards
VB~10 (M8; \citealt{bie61,kir91}) and 
2MASS~J14392836+1929149 (L1, hereafter 2MASS~J1439+1929; \citealt{kir99}).
Despite the poor observing conditions, the data for {\namesh} have exceptionally
good signal-to-noise, $\gtrsim$150 in the $JHK$ flux peaks and $\sim$50 in the
bottom of the 1.4 and 1.8~$\micron$ {\wat} bands.  Color biases due to 
telluric cloud absorption do not appear to be present, as indicated by comparison of 2MASS photometry and 
synthetic $J-H$, $H-K_s$ and $J-K_s$ colors computed from the spectral data, which agree
to within the photometric uncertainties.

The morphology of the near-infrared spectrum of {\namesh} 
is typical of a late-type M
or early-type L dwarf, with bands of TiO and VO absorption at red optical
wavelengths ($\lambda < 1$~{\micron}); prominent {\wat} absorption
at 1.4 and 1.8~$\micron$; FeH absorption at 0.99,
1.2 and 1.55~$\micron$;
{\nai} and {\ki} line absorption in the 1.0-1.3~$\micron$ region;
weak {\nai} lines at 2.2~$\micron$; and strong CO bandheads 
at 2.3--2.4~$\micron$.  For the most part, the spectrum of
{\namesh} is more consistent with that of 2MASS~J1439+1929; note
in particular the similarities in the overall shape of the 1.0--1.35~$\micron$ $J$-band
flux peak and the deep 1.4~$\micron$ {\wat} band.
However, TiO and VO bands are more similar
to (but weaker than) those seen in the spectrum of VB~10, 
while the weak 2.2~$\micron$ {\nai} lines are rarely seen in L dwarf
spectra (e.g., \citealt{mcl03}).  The near-infrared spectrum
of {\namesh} is also somewhat bluer than that of 2MASS~J1439+1929, 
in line with their respective colors ($J-K_s$ = 1.13$\pm$0.04 versus
1.21$\pm$0.03). 

The similarities to 2MASS~J1439+1929  
suggests an L1 near-infrared spectral type
for {\namesh}, which is confirmed by examination of
the spectral indices and index/spectral type relations
defined by \citet{rei01a} and \citet{geb02}.
The average subtype for the four indices K1 (measuring the shape of the $K$-band
flux peak; \citealt{tok99}), {\wat}-A, {\wat}-B and {\wat}-1.5 (all measuring the strength of the
1.4~$\micron$ {\wat} band)
yields a near-infrared classification of L1 ($\pm$0.6 subtypes), 
consistent with the L0.5 
near-infrared classification reported by \citet{wil03}.

This classification is fully three subtypes later
than the M8: optical spectral type reported by \citet{cru03}.  
However, such discrepancies are
not altogether uncommon amongst late-type dwarfs.
\citet{geb02} and \citet{kna04} have
found disagreements of up to 1.5~subtypes
between optical (based on the \citealt{kir99} scheme)
and near-infrared classifications (based on their own scheme)
for several L dwarfs. 
\citet{me1126} have discussed a subclass of unusually blue L
dwarfs whose optical classifications are consistently 2-3 subtypes
earlier than their near-infrared classifications.
Such discrepancies have been variously attributed to 
surface gravity, metallicity, condensate cloud or multiplicity effects 
(e.g., \citealt{kna04,chi06,cru07,fol07,me1126}).
The large {\vtan} of {\namesh}, indicating that this source may be somewhat older,
suggests that high surface gravity and/or slightly subsolar metallicity
could explain the discrepant optical and near-infrared spectral classifications.

However, {\namesh} exhibits one unusual feature not seen in 
the comparison spectra in Figure~\ref{fig_nirspec}, a slight dip at 1.6~$\micron$,
that  suggests multiplicity is relevant in this case.
The 1.6~$\micron$ feature is nearly coincident with the
1.57-1.64~$\micron$ FeH absorption band
commonly observed in L dwarf near-infrared 
spectra \citep{wal01,cus03}.  Yet its morphology is clearly different,
with a cup-shaped depression as opposed to the flat plateau
seen in the comparison spectra of Figure~\ref{fig_nirspec}.
More importantly, this feature has the same morphology and is centered at the same
wavelength as the peculiar feature noted in the spectrum of 
SDSS~J0805+4812 \citep{me0805}.  In that case, the
1.6~$\micron$ feature and other spectral peculiarities
were attributed to the presence of  
a mid-type T dwarf companion.  Given the similar discrepancy in optical
and near-infrared classifications for SDSS~J0805+4812 (L4 and L9.5, respectively),
it is reasonable to consider whether {\namesh} also harbors a faint T dwarf
companion.

\subsection{High Angular Resolution Imaging}

In an attempt to search for faint companions,
{\namesh} was imaged on 2008~January~15 (UT) 
with the sodium LGS AO system \citep{wiz06,van06} and
facility near-infrared camera NIRC2 on the 10m Keck Telescope.
Conditions were photometric with average/below-average
seeing.  The narrow field-of-view camera of NIRC2 was utilized, 
providing an image scale of
$9.963\pm0.011$~mas/pixel \citep{pra06} over a
$10.2\arcsec \times 10.2\arcsec$ field of view.  
All observations were conducted using the 
MKO\footnote{Mauna Kea Observatory (MKO) photometric system; \citet{sim02,tok02}.} 
$K_s$-band filter.
The LGS provided the
wavefront reference source for AO correction, while 
tip-tilt aberrations and quasi-static changes were measured
contemporaneously by
monitoring the $R=16.7$~mag field star USNO-B1.0~0852$-$0031783
\citep{mon03}, located 14\arcsec\ away from {\namesh}.
The LGS, with an equivalent  
brightness of a $V\approx10.4$ mag star, 
was pointed at the center of the NIRC2
field-of-view for all observations.

{\namesh} was imaged using two different methods in order to probe
the widest possible range of projected separations: (1)~direct imaging
and (2)~aperture mask interferometry.  
In the first case, a series of 3 dithered 60-second images was obtained,
offsetting the telescope by a few arcseconds between exposures, 
for a total integration time of 180~seconds.  
Raw frames were reduced using standard
procedures.  Normalized flat field frames were constructed from the differences of images of
the telescope dome interior with and without continuum lamp
illumination.  A master sky frame was created from the median
average of the bias-subtracted, flat-fielded images and subtracted
from the individual exposures.  Individual frames were registered and stacked to
form a final mosaic imaged.  The observations achieved a 
point spread function (PSF) full-width at half-maximum of
 0{\farcs}07 and a Strehl ratio of 0.21.  
With the exception of the primary target, no sources were detected in
a $6\arcsec \times 6\arcsec$ region centered on {\namesh}.

Aperture mask observations were also obtained with the LGS AO+NIRC2
instrumental setup. In this method, a 9-hole aperture mask is placed in a
filter wheel near a re-imaged pupil plane within the NIRC2 camera. The
mask has non-redundant spacing, so each Fourier component of
the recorded image corresponds to a unique pair of patches on the Keck
primary mirror. The primary interferometric
observables of squared visibility and closure-phase are therefore calibrated much
better than images using the full aperture. This technique has a
long history of achieving the full diffraction limit of a telescope
(e.g. \citealt{mic20,bal86,nak89,tut00}) and has been recently
applied to natural guide star AO  
observations at Keck \citep{ire08,kra08}.
This is the first application of aperture mask interferometry to LGS AO observations
that we are aware of.

{\namesh} was observed in this setup
using a two-point dither pattern, with five 50-second
integrations at each dither position.  
The nearby field star 2MASS~J03381363$-$0332508, which has a
similar $K_s$-band brightness and tip-tilt star asterism as {\namesh},
was contemporaneously observed to calibrate both
instrumental closure phase and visibility.
Images of the interferograms formed by the mask were recorded
by the NIRC2 detector, and squared visibilities and closure-phases were
extracted from the image Fourier transforms.
Raw visibility amplitudes were $\sim$0.05 on the longest baselines.
The closure phases for this calibrator star
were subtracted from those of {\namesh},
while the calibrator's squared visibilities were divided into 
those of {\namesh}.
The one-sigma scatter in the  calibrated closure phase was 5$\degr$.
Using standard analysis techniques (e.g., \citealt{kra08}), we found no evidence of a
binary solution in the data. 

Upper limits on the presence of a faint companion to {\namesh}
were computed separately for the direct imaging and aperture
mask observations.  For the direct imaging data, upper limits 
were determined by first smoothing
the final mosaic with an analytical representation of the PSF's radial
profile, modeled as the sum of multiple gaussians.  We then measured
the standard deviation of flux counts in concentric annuli out to
3$\arcsec$ in radius centered on the science target, normalized by the
peak flux of the science target. We considered
10$\times$ these values as the
flux ratio limits for any companions, as
visually verified by inserting fake sources into the
image using translated and scaled versions of the science target.
For the aperture mask data, detection limits at 99\% confidence were
calculated in three annuli spanning 0$\farcs$020--0$\farcs$16 
in separation (the lower limit corresponding to the diffraction limit of
the aperture mask) using the Monte-Carlo approach described in \citet{kra08}.

\begin{figure}
\epsscale{0.8}
\plotone{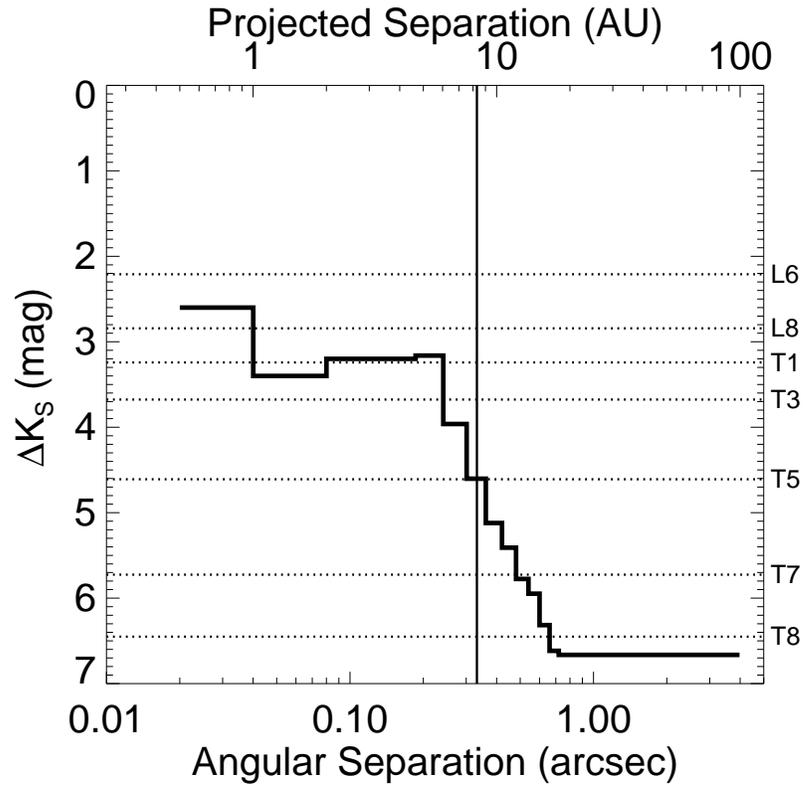}
\caption{Upper limits on the relative $K_s$-band flux ratio of a 
faint companion to {\namesh} as a function of separation based on LGS AO
observations.  The limits shortward of 0$\farcs$25 are based on observations
with a 9-hole, non-redundant aperture mask, while those longward are based on direct imaging
observations.   Angular separation in arcseconds is mapped
onto projected separation in AU at the estimated distance of 25 pc.
Flux ratios are mapped onto secondary spectral type using the 
``bright'' MKO $M_K$/spectral type relation of \citet{liu06} and spectral type-dependent
filter transformations from \citet{ste04}. 
\label{fig_ao}}
\end{figure}

Figure~\ref{fig_ao} displays the resulting flux ratio limits for a faint companion
as a function of separation for both datasets.  At separations $\lesssim$0$\farcs$25,
the aperture mask data exclude any companions with $\Delta{K_s} \lesssim$ 
3~mag for separations down to 0$\farcs$04.  Note that better seeing, as opposed
to longer integrations, would have provided greater improvement in sensitivity
in this range.  The direct imaging observations exclude any companions
with $\Delta{K_s} \lesssim$ 7~mag at separations ${\gtrsim}0{\farcs}7$,
with the floor set primarily by sky shot noise and detector read noise.
These limits are discussed further in $\S$~4.1.

\section{Binary Template Matching}

\subsection{Spectral Sample}

As an alternative method to identify and characterize a possible companion to {\namesh},
a variant of the binary spectral template matching technique described in 
\citet{me0805} was applied to the near-infrared spectral data.\footnote{See 
also \citet{me0423,mehst2,me1126,rei2252,meltbinary,loo07,sie07}; and \citet{loo08}.}
In this method, the spectrum of a late-type source is compared to a large set
of binary spectral templates constructed from empirical data for
M, L and T dwarfs.  The component spectra of each binary template 
were scaled according to
empirical absolute magnitude/spectral type relations. 
To minimize systematic effects, source and template spectra
are required to have the same resolution and wavelength coverage,
which is facilitated in this case by using a  sample of
nearly 200 SpeX prism spectra of M5--T8 dwarfs drawn from 
the literature\footnote{See 
\citet{mewide3,meclass2,me1520,cru04,sie05,metgrav,me2200,chi06,mce06,rei2252,mehd3651,meltbinary,lie07,loo07}; and \cite{luh07}.  These data are available at \url{http://www.browndwarfs.org/spexprism}.} and our own unpublished
observations.  
Spectral types for the sources in this sample were assigned according to 
published classifications,\footnote{A current list of L and T dwarfs with their
published optical and near-infrared spectral types is maintained by C.\ Gelino, J.\ D.\
Kirkpatrick and A.\ Burgasser at \url{http://www.dwarfarchives.org}.}
based either on the optical classification schemes of \citet{kir91} and \citet{kir99} for M5--L8 dwarfs or the near-infrared classification scheme
of \citet{meclass2} for L9--T8 dwarfs (M and L dwarfs with only near-infrared
classifications reported were not included here).
The initial spectral sample was purged of low signal-to-noise data as well as 
spectra of those sources known to be binary or noted as 
peculiar in the literature (e.g., 
low surface gravity brown dwarfs, 
subdwarfs, etc.).
This left a sample of 132 spectra of 125 sources, 
listed in Table~\ref{tab_templates}.  

\subsection{Single Template Fits}

To ascertain whether an unresolved binary truly provides a better
fit to the spectrum of {\namesh}, comparisons were first made to 
individual sources in the SpeX sample.
All spectra were initially
normalized to their peak flux in the 1.2--1.3~$\micron$ band.  
The statistic $\sigma^2$ was then computed between the {\namesh} 
($f_{\lambda}(0320)$) and template 
spectra ($f_{\lambda}(T)$), where 
\begin{equation}
\sigma^2 \equiv \sum_{\{ \lambda\} }\frac{[f_{\lambda}(0320)-f_{\lambda}(T)]^2}{f_{\lambda}(0320)}
\end{equation}
(see \citealt{me0805}). The summation is  performed over the 
wavelength ranges $\{\lambda\}$ =
0.95--1.35~$\micron$, 1.45--1.8~$\micron$ and 2.0--2.35~$\micron$ in order
to avoid regions of strong telluric absorption. The denominator 
provides a rough estimate of shot noise in the spectral data, which is dominant
in the highest signal-to-noise spectra, and therefore makes $\sigma^2$ a rough
approximation of the $\chi^2$ statistic.\footnote{In the near-infrared, foreground emission
generally dominates noise contributions. However, given the broad range of observing conditions
in which the {\namesh} and template data were taken, we chose not to include this 
term in our $\sigma^2$ statistic.} 
To eliminate normalization biases, each template spectrum was additionally
scaled by a multiplicative factor in the
range 0.5--1.5 to minimize $\sigma^2$.  

\begin{figure}
\epsscale{1.1}
\plottwo{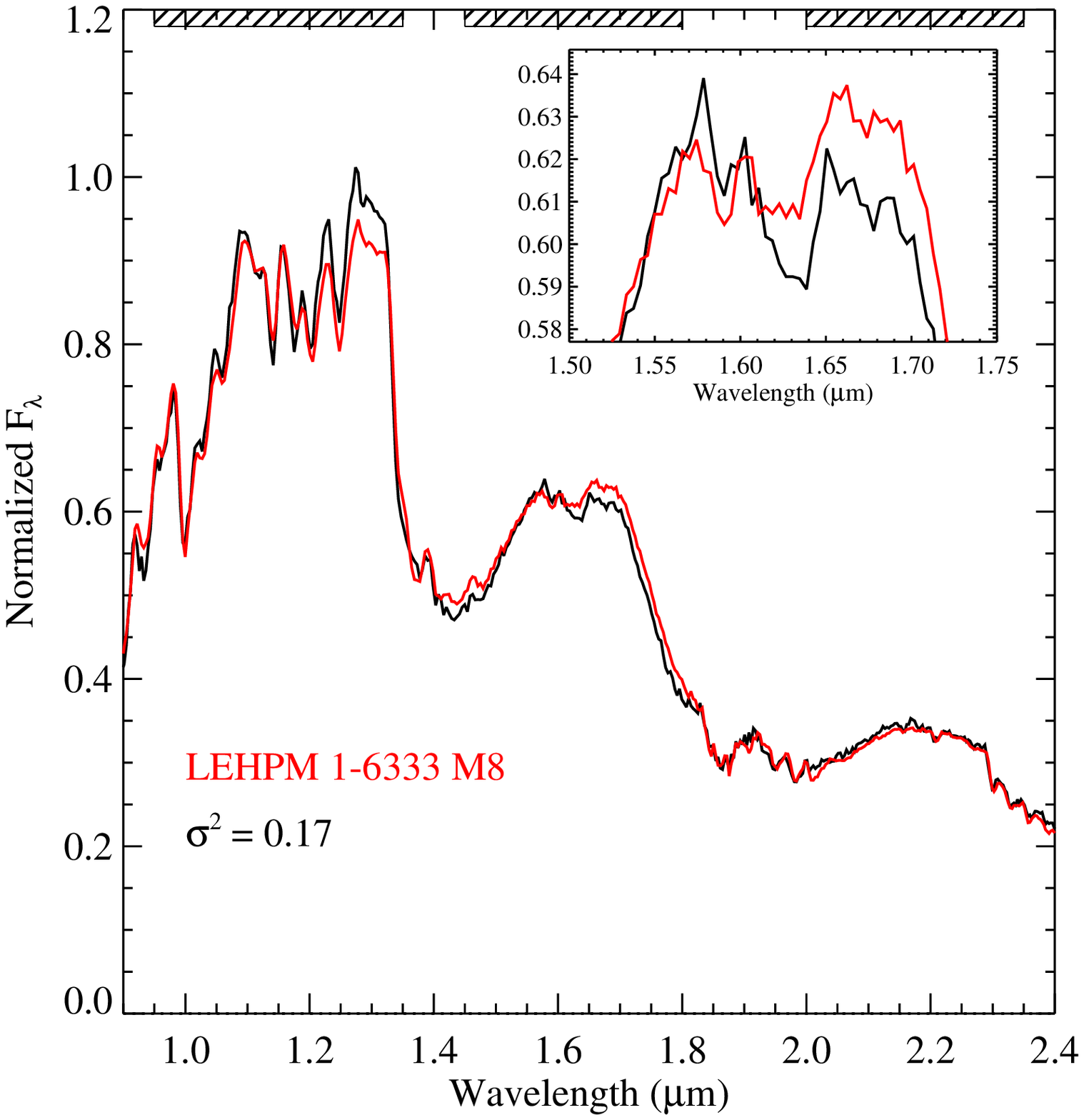}{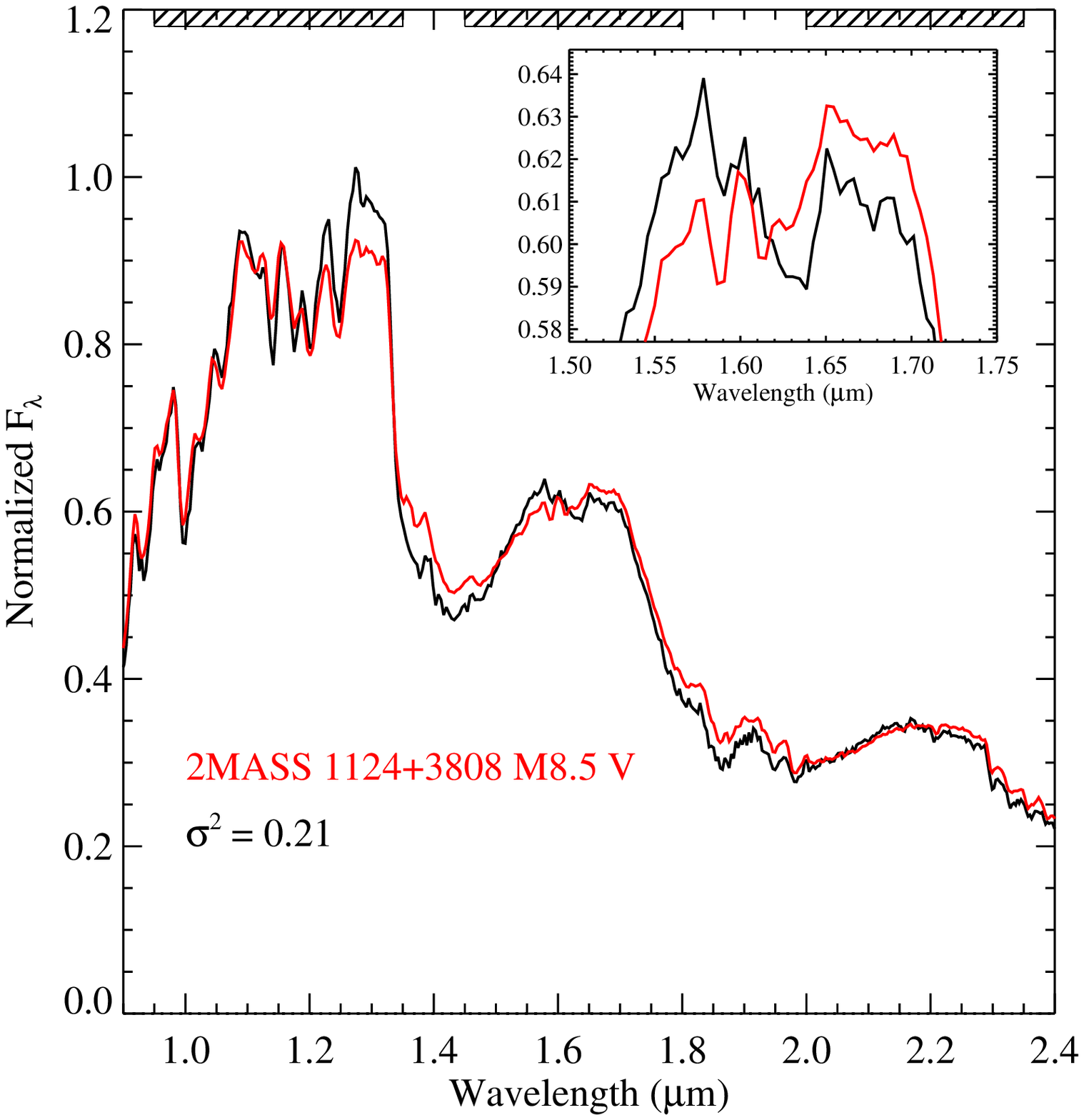}
\plottwo{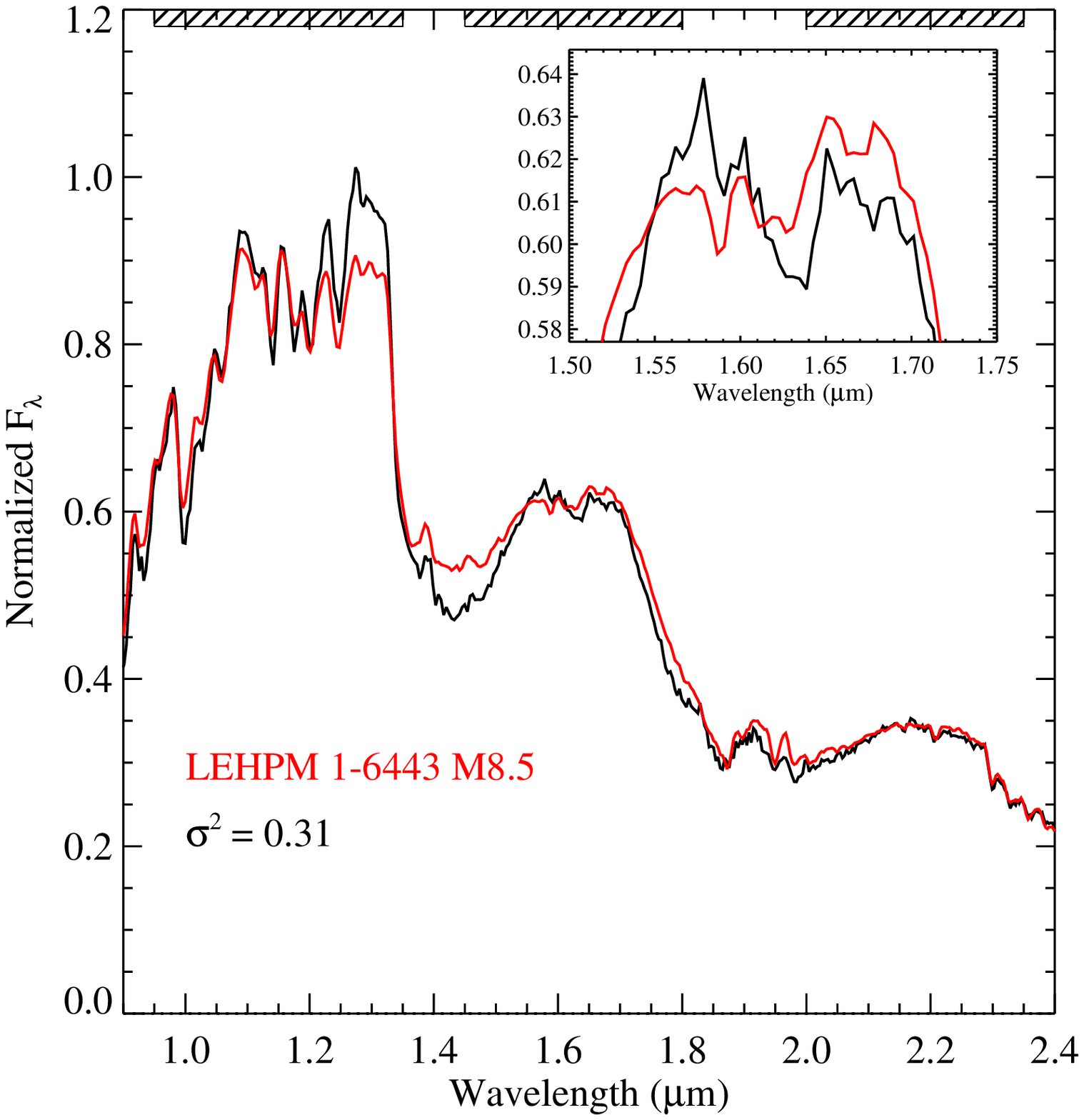}{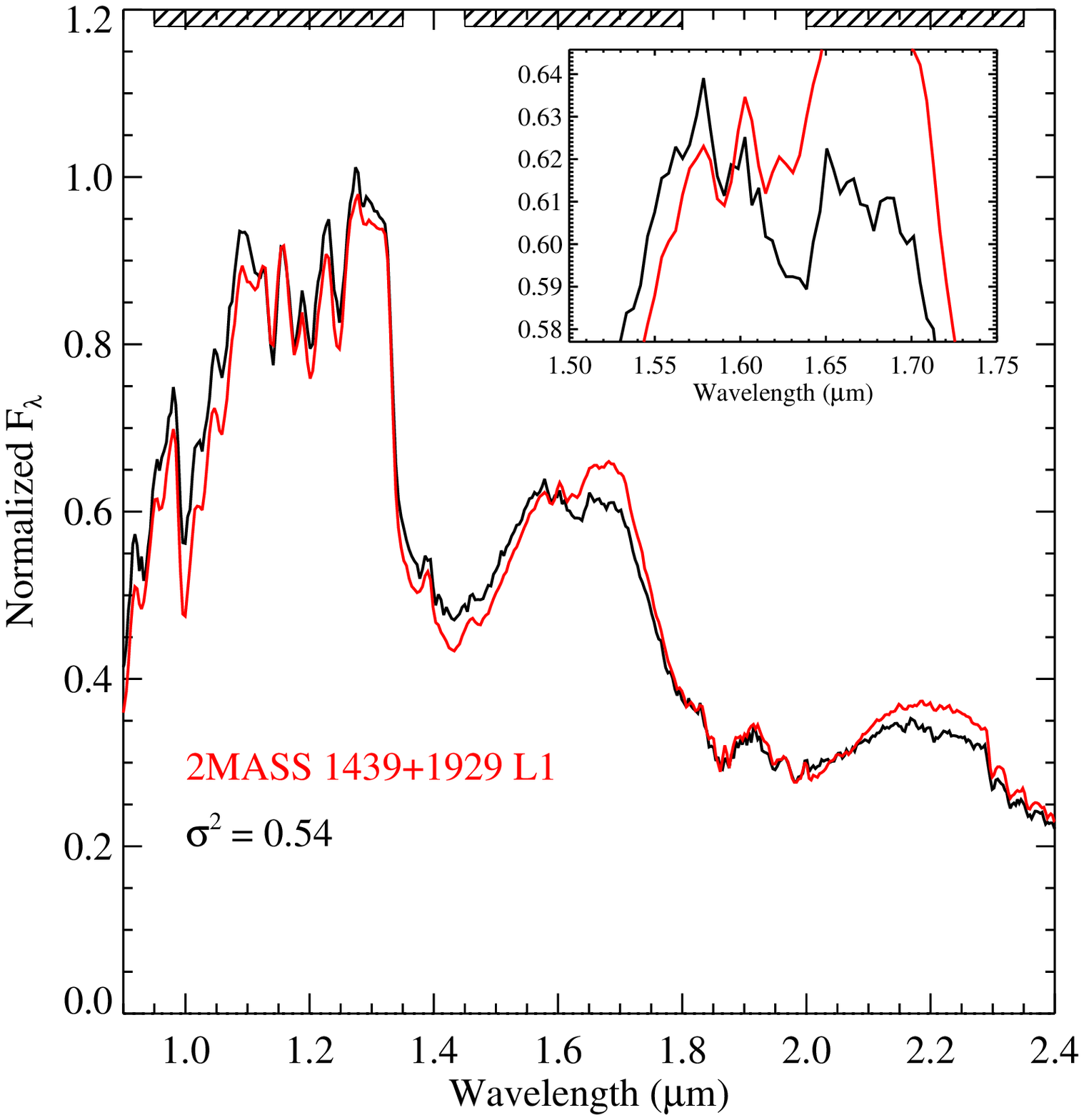}
\caption{Best fit single spectral templates 
(red lines) to the spectrum of {\namesh} (black lines):
LEHPM 1-6333 (M8, $\sigma^2$ = 0.17),
2MASS~J1124+3808 (M8.5, $\sigma^2$ = 0.21),
LEHPM 1-6443 (M8.5, $\sigma^2$ = 0.31) and
2MASS~J1439+1929 (L1, $\sigma^2$ = 0.54).
All spectra are normalized in the 1.2--1.3~$\micron$
window, with the templates further scaled to minimize
their $\sigma^2$ deviations.  The spectral
bands used to calculate $\sigma^2$ are indicated at the top of each panel.
Inset boxes show a close-up of the 1.5--1.75~$\micron$ region
where the peculiar 1.6~$\micron$ feature present in the spectrum
of {\namesh} is located.
\label{fig_single}}
\end{figure}

Figure~\ref{fig_single} displays the four best
single template matches, all having $\sigma^2 < 0.6$.
The three best-fitting sources---LEHPM~1-6333 (M8),
2MASS~J1124+3808 (M8.5) and LEHPM~1-6443 (M8.5)---have
optical spectral types consistent with the 
optical type of {\namesh}.  The fourth-best fit, the L1 2MASS~J1493+1929, 
was shown to provide an adequate
match to the spectrum of {\namesh} in Figure~\ref{fig_nirspec}.
The LEHPM\footnote{Liverpool-Edinburgh High Proper Motion (LEHPM) Catalog of \citet{por04}.} sources
have large proper motions ($\mu > 0\farcs$4~yr$^{-1}$), notably
similar to {\namesh}.
All four sources shown in Figure~\ref{fig_single}
provide reasonably good matches
to the broad near-infrared spectral energy distribution of {\namesh},
but with two key discrepancies: an absence of the 1.6~$\micron$ feature
(inset boxes in Figure~\ref{fig_single}) and a shortfall in the peak
spectral flux at 1.27~$\micron$.  In the first case, FeH absorption bands 
are clearly seen in the comparison spectra
but do not produce the distinct dip seen in the spectrum of {\namesh}.
In the second case, the spectrum
of {\namesh} is consistently brighter in the 1.2--1.35~$\micron$
range as compared to the (appropriately scaled) late-type M dwarf templates.  
As demonstrated below, both of these 
discrepancies can be resolved by the addition of a T dwarf component.

\subsection{Binary Template Fits}

Binary spectral templates from the SpeX prism sample
were constructed by first
flux-calibrating each spectrum according to established
absolute magnitude/spectral type relations.  For M5-L5 dwarfs,
the 2MASS $M_J$/spectral type relation of \citet{cru03} was used.
For L5-T8 dwarfs, both of the 
MKO $M_K$/spectral type relations defined in \citet{liu06}
were considered.
The Liu et al.\ relations are based on a sample of L and T dwarfs with measured
pallaxes and MKO photometry, but one relation (``bright'') was constructed after
rejecting known (resolved) binaries while the other relation
(``faint'') was constructed after rejecting
all known and {\em candidate} binaries as described in that study.
As illustrated in Figure~3 of \citet{meltbinary}, these two relations
envelope the $M_K$ values of currently measured sources (including
components of resolved binaries), but diverge by as much as $\sim$1~mag
for spectral types L8--T5.  Nevertheless, the \citet{liu06} relations represent our current
best constraints on the absolute magnitude/spectral type relation across the L dwarf/T dwarf transition.
In all cases, 
synthetic magnitudes to scale the data were calculated directly from the spectra.
Binary templates were then constructed by adding together
the calibrated spectra of source pairs
whose types differ by at least 0.5 subclasses, producing a total of 8248 
unique combinations.  The binary templates were then normalized
to their peak flux in the 1.2--1.3~$\micron$ band and compared
to the spectrum of {\namesh} in the same manner as the single
source templates; i.e., with additional scaling to minimize $\sigma^2$. 

\begin{figure}
\epsscale{1.1}
\plottwo{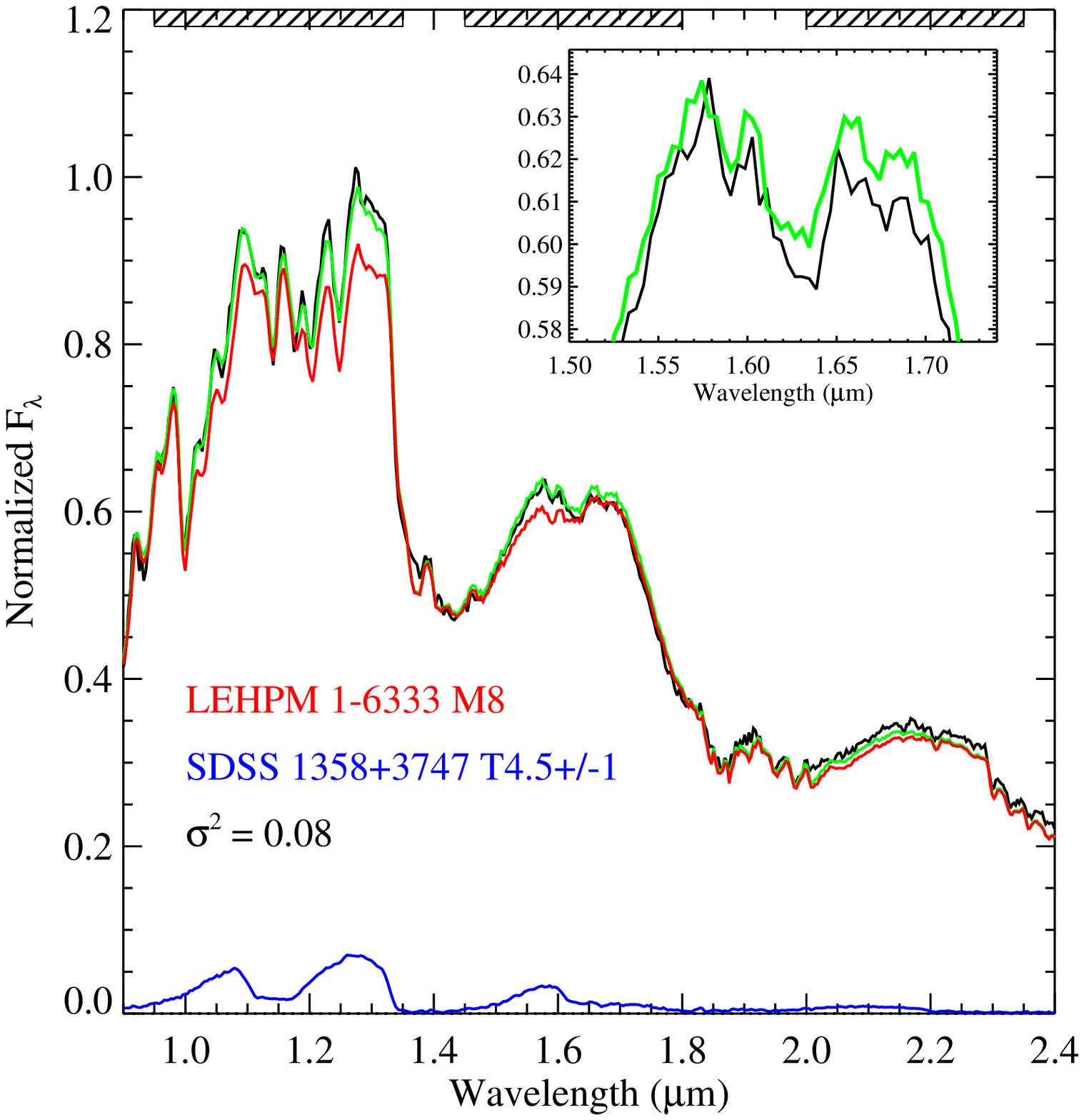}{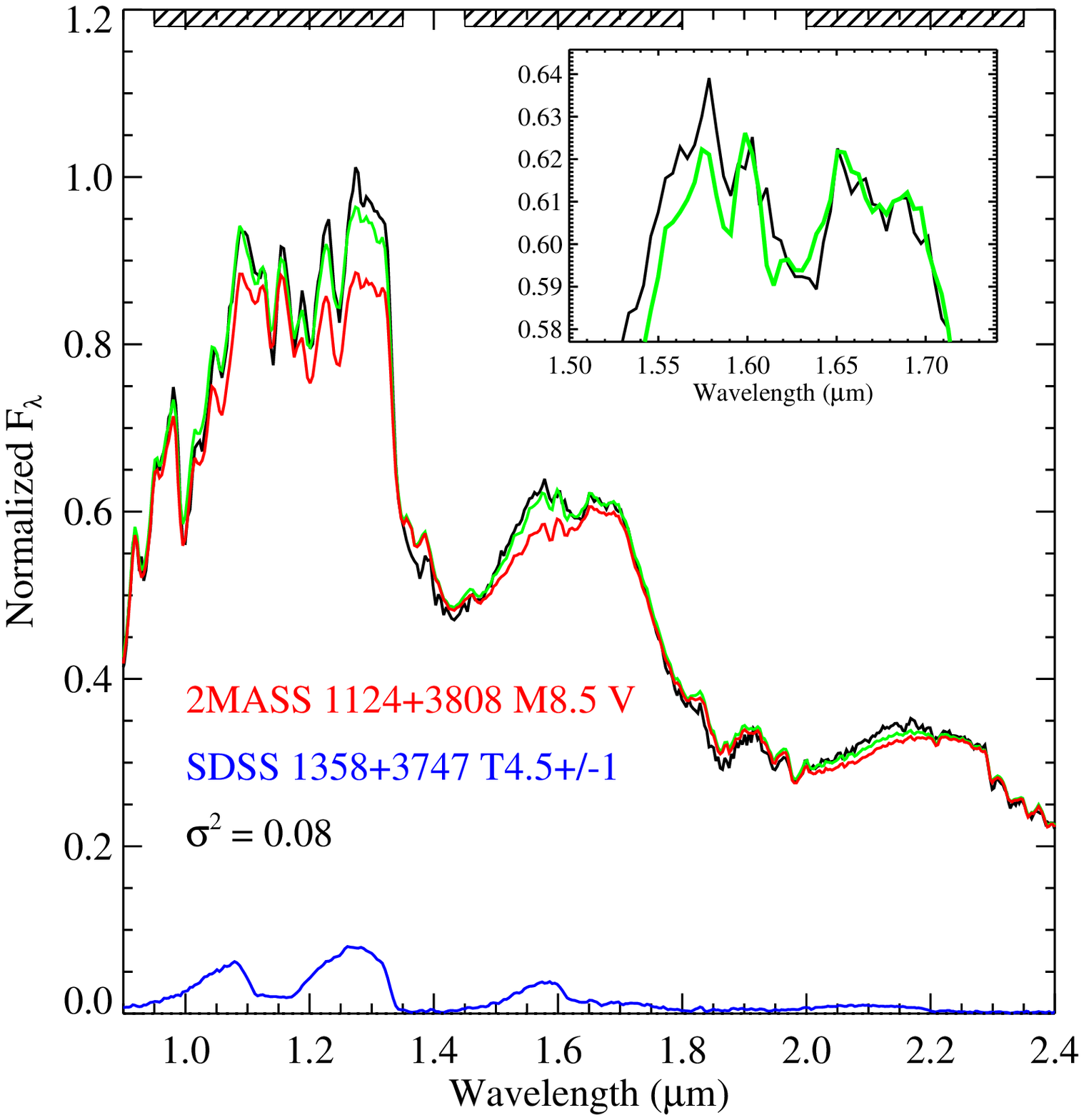}
\plottwo{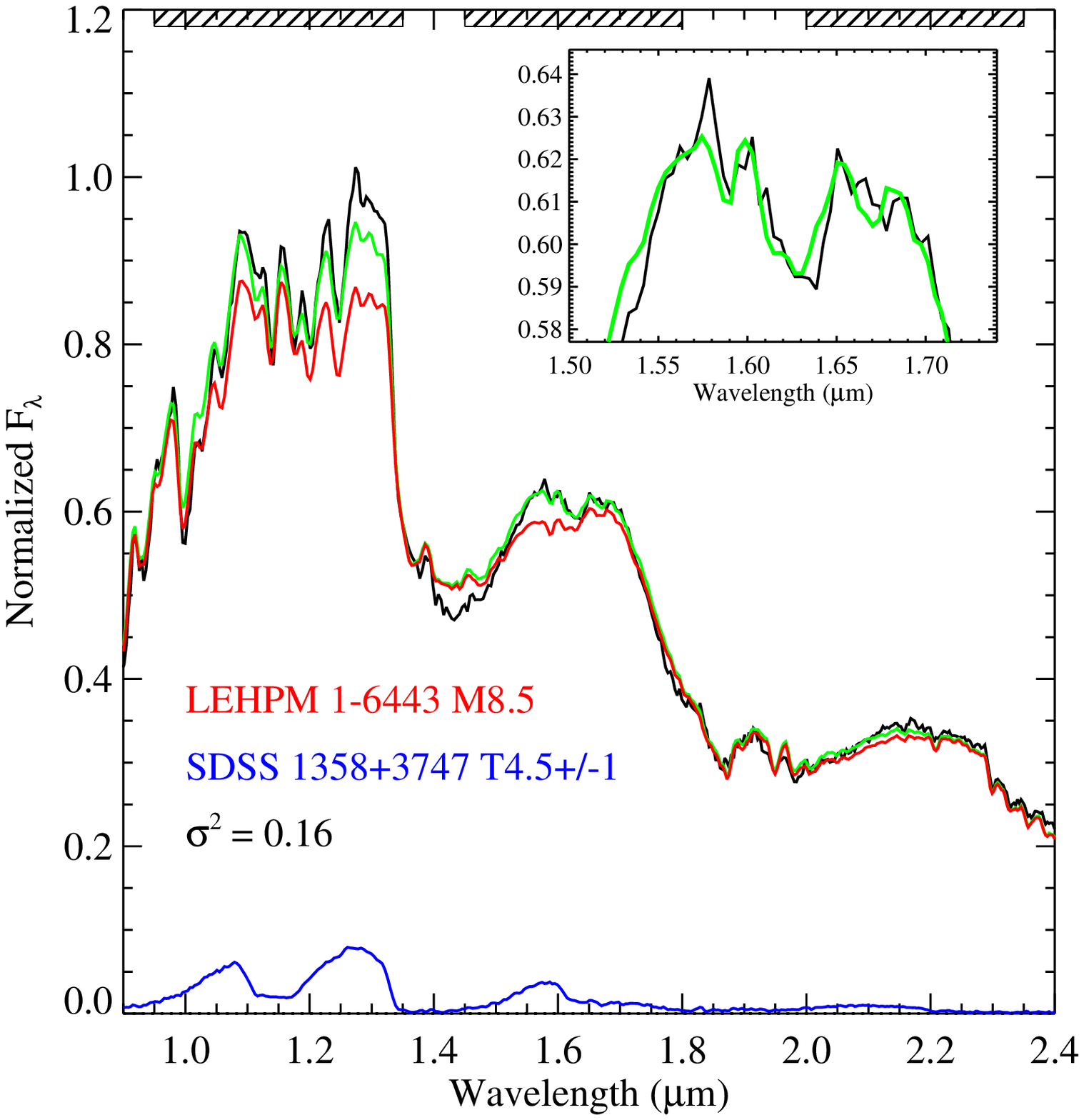}{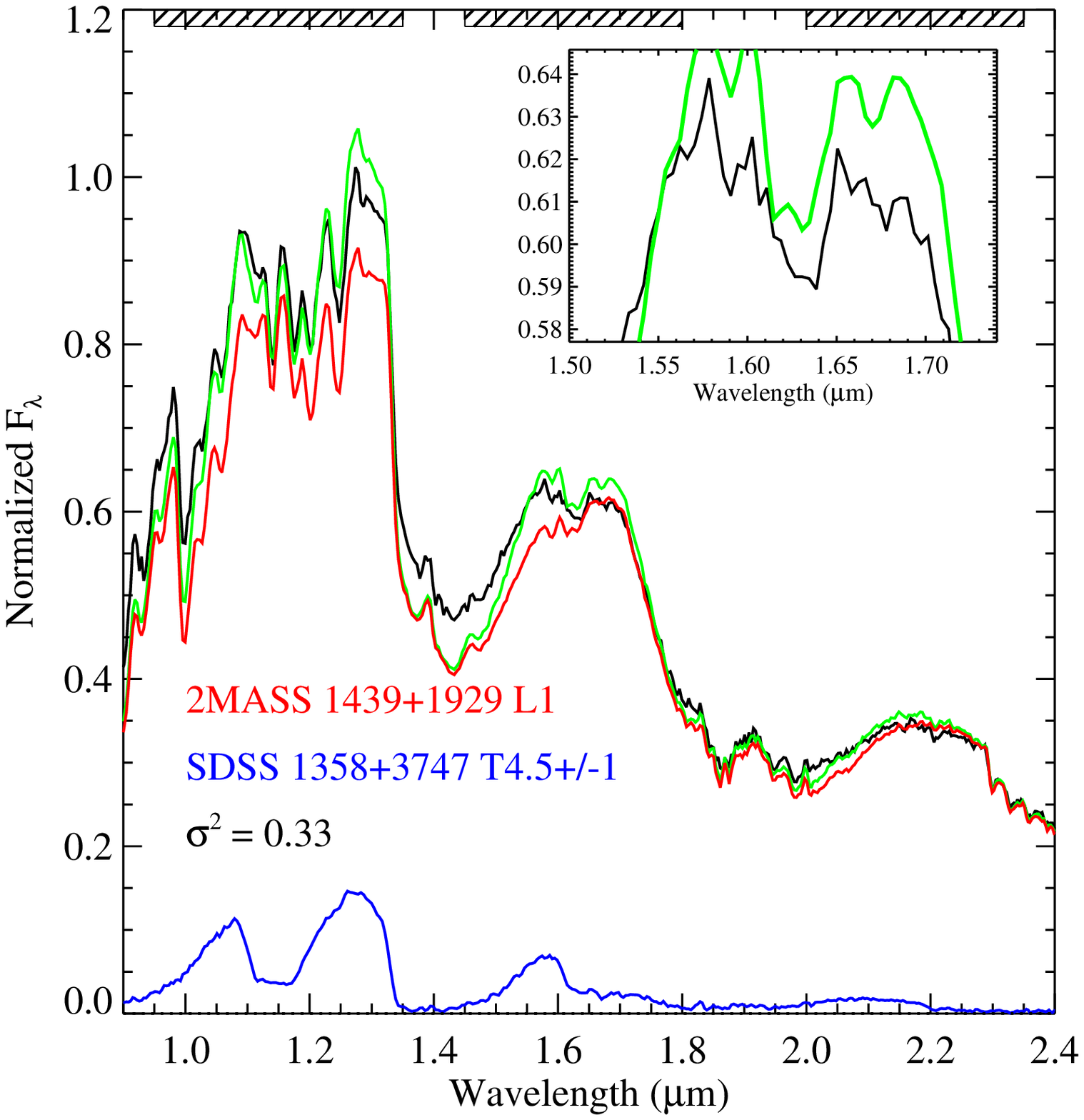}
\caption{Best fit binary spectral templates 
(green lines) to the spectrum of {\namesh} (black lines) 
constructed from the primaries shown in Figure~\ref{fig_single}
and using the ``faint'' $M_K$/spectral type
relation of \citet{liu06}.  The primary
(red lines) and secondary (blue lines) component spectra
are shown scaled in accordance with their contribution to the
binary templates.  Inset boxes show a close-up of the 
1.5--1.75~$\micron$ spectra of {\namesh} and binary templates.
\label{fig_double}}
\end{figure}

Figure~\ref{fig_double} displays the best fitting binary templates
constructed from the primaries shown in Figure~\ref{fig_single}
and using the ``faint'' $M_K$/spectral type relation of
\citet{liu06}.  
For all four cases, the addition of a
mid-type T dwarf secondary spectrum considerably improves
the spectral template match.  In particular, the 1.6~$\micron$
spectral dip is very well reproduced, 
while the flux peaks at 1.27~$\micron$ in the binary templates
are more consistent with the spectrum of {\namesh}.
Even detailed alkali line and FeH features in the 0.9--1.3~$\micron$ region
are better matched with the binary templates.  

\begin{figure}
\epsscale{1.1}
\plottwo{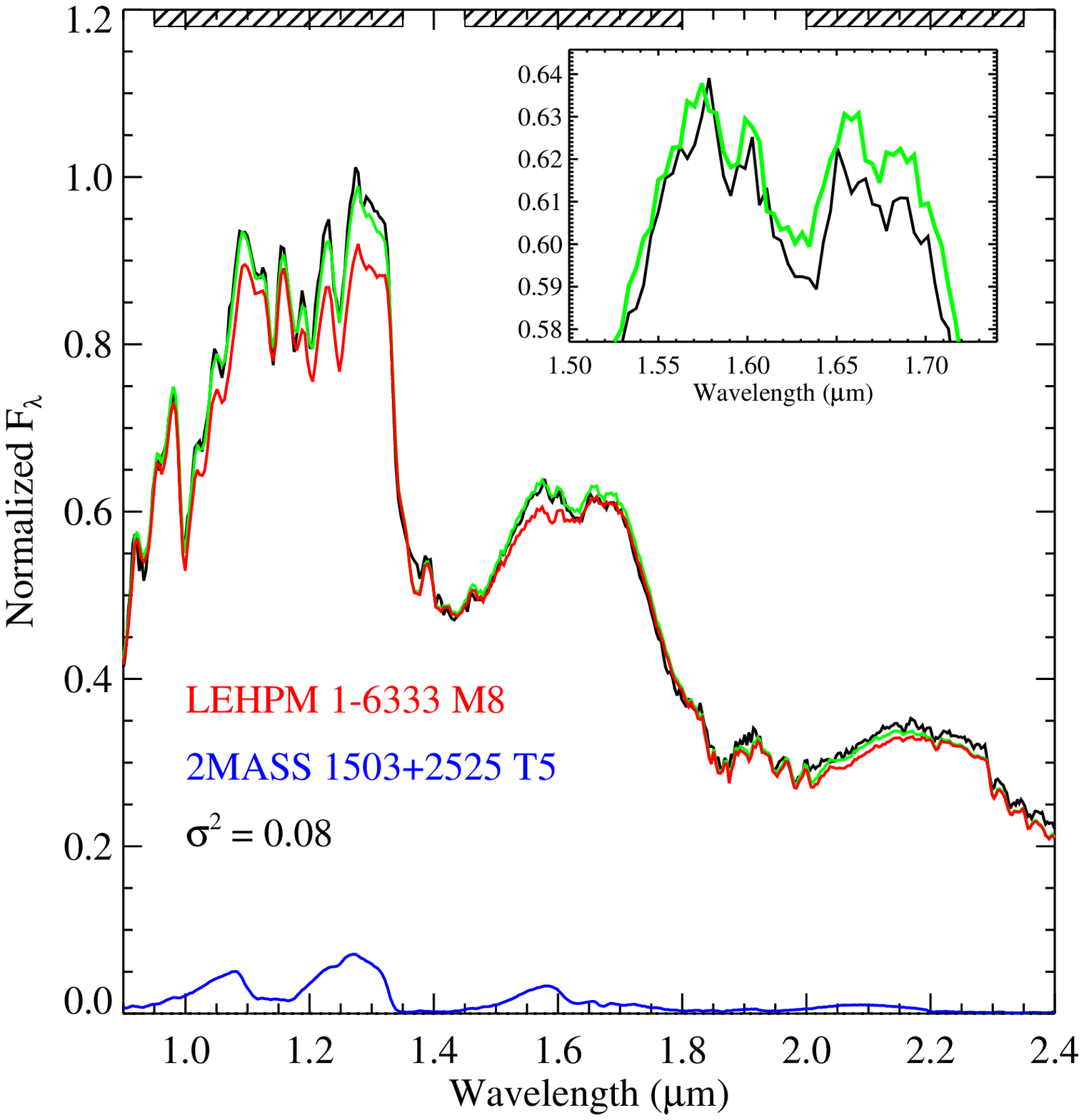}{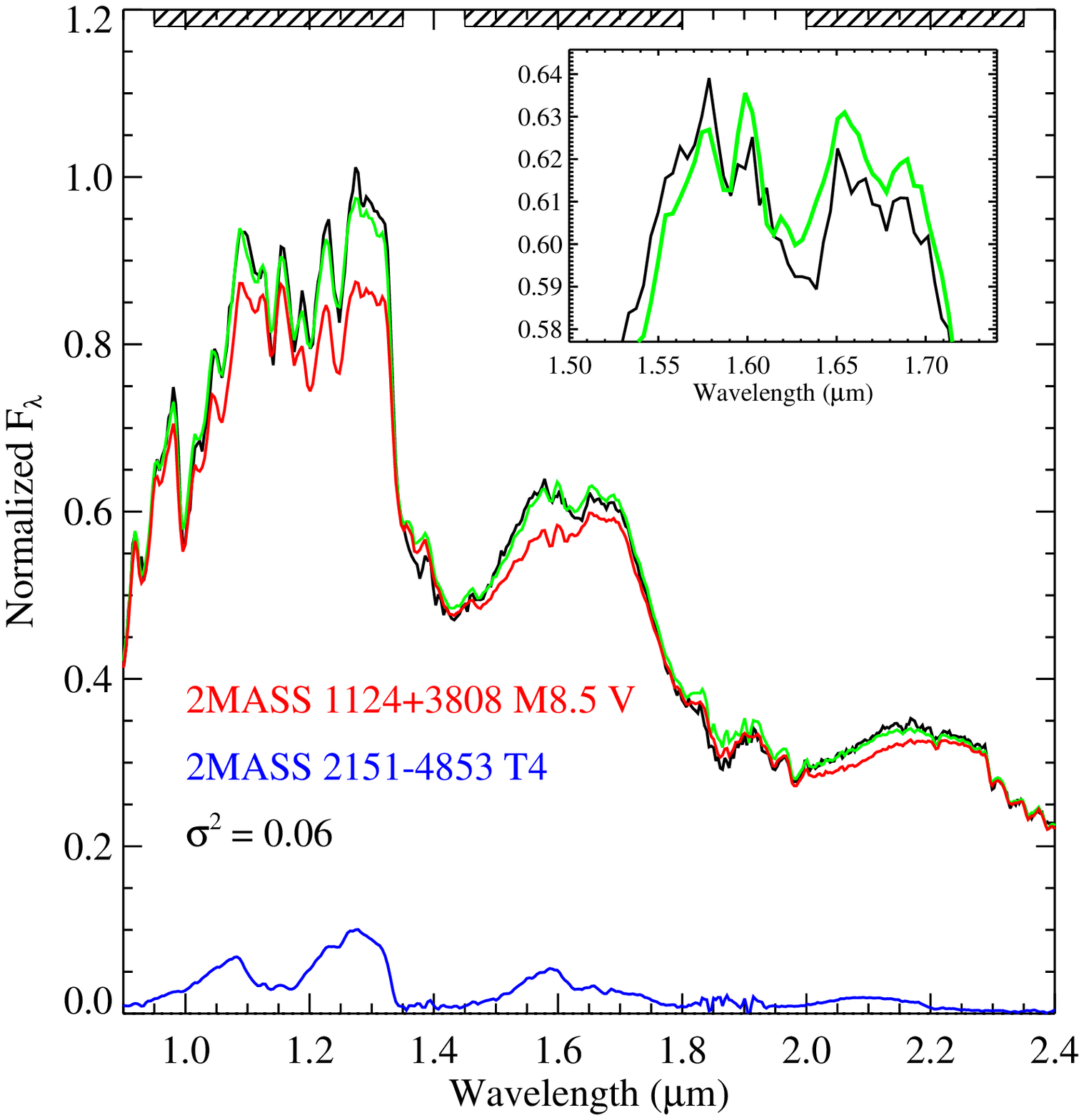}
\plottwo{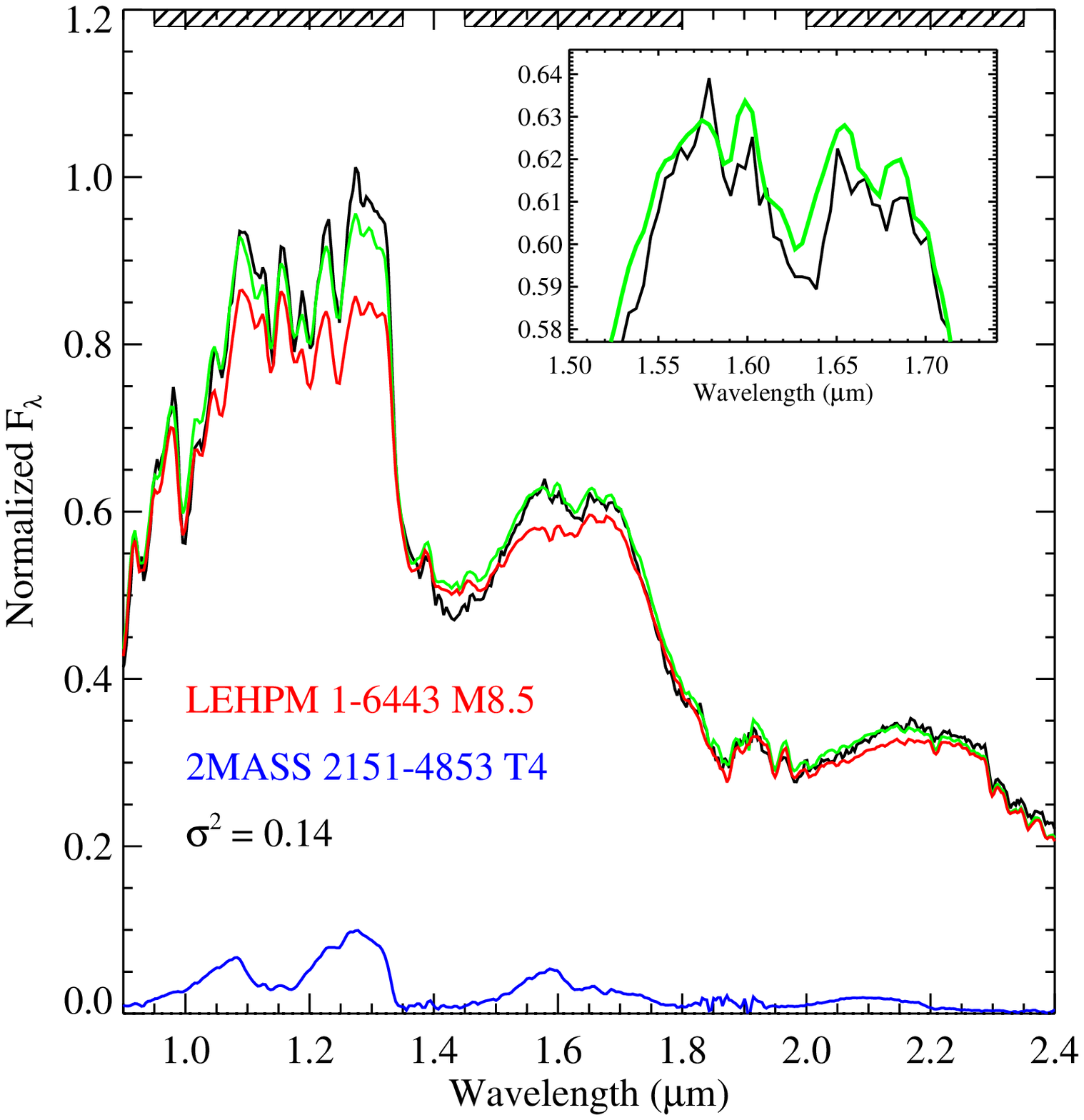}{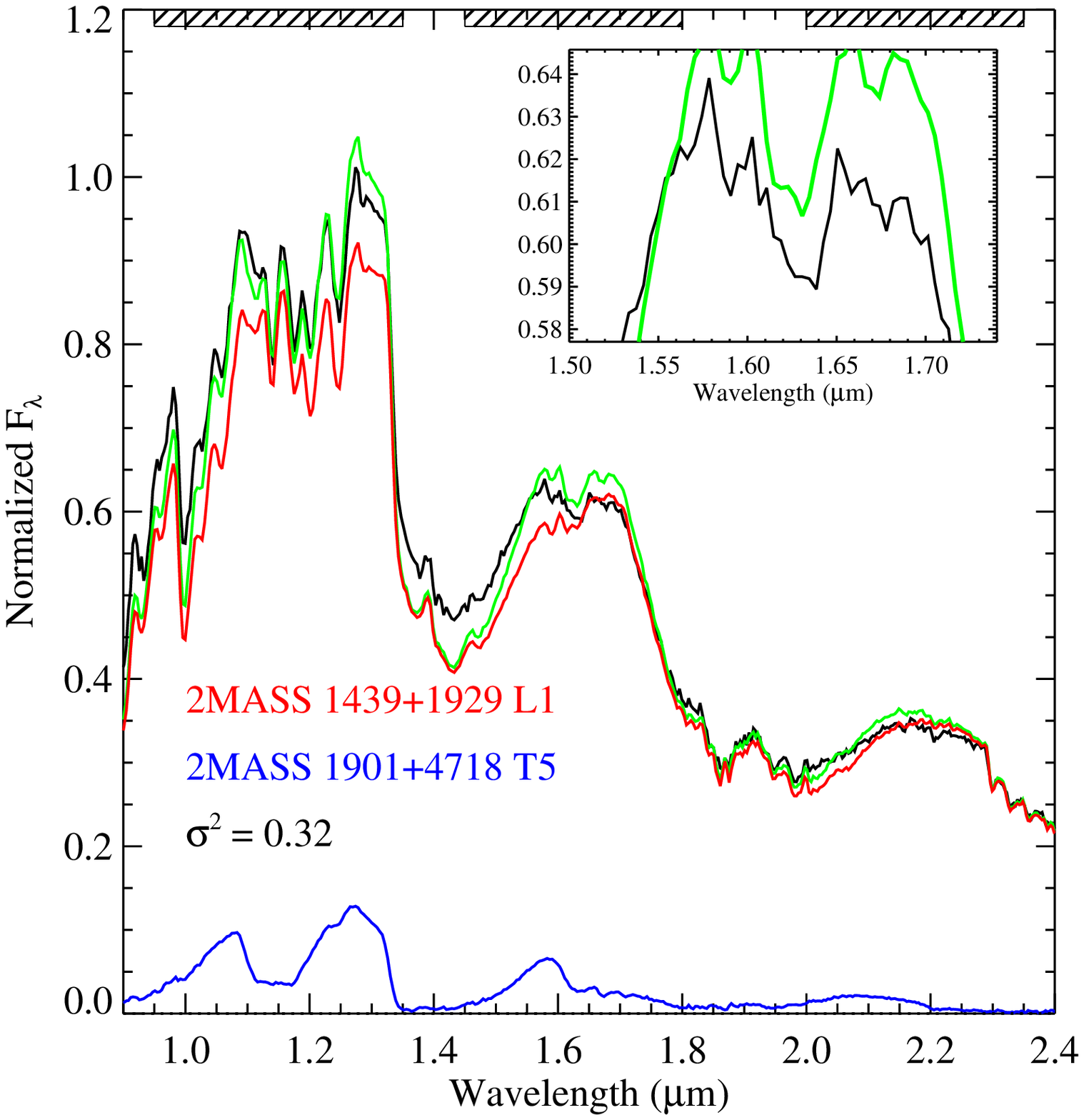}
\caption{Same as Figure~\ref{fig_double2} but based on binary spectral templates 
constructed using the ``bright'' $M_K$/spectral type
relation of \citet{liu06}.
\label{fig_double2}}
\end{figure}

Figure~\ref{fig_double2} displays the best fitting 
binary templates using the ``bright''
$M_K$/spectral type relation of
\citet{liu06}.  
There is a small degree of improvement in these fits over
those using the ``faint'' $M_K$ relation, although the
differences are very subtle due to the very small contribution
of light by the T dwarf secondaries 
($\Delta{J} \approx$ 3.5~mag, depending on the components).
This result is fortuitous, as it indicates that the better fits provided
by the binary templates are only weakly dependent on the absolute
magnitude relation assumed over a spectral
type range in which such relations are currently most uncertain.

Besides the best-fit comparisons shown in Figures~\ref{fig_double} 
and~\ref{fig_double2}, there were many excellent matches ($\sigma^2 < 0.1$)
found among binaries templates which had LEHPM 1-6333 or
2MASS~J1124+3808 as primaries: 30 for the ``faint'' $M_K$/spectral type
relation and 58 for the ``bright'' relation.  
The average primary and secondary spectral types
for the combinations in this well-matched sample are 
M8.5$\pm$0.3 and T5.0$\pm$0.9, respectively,
with no significant differences between analyses using the
``faint'' or ``bright'' $M_K$/spectral type relations.  The mean relative
magnitudes of the primary and secondary components were 
$\Delta{J} = 3.5{\pm}0.2$~mag, $\Delta{H} = 4.3{\pm}0.3$~mag, 
$\Delta{K} = 4.9{\pm}0.3$~mag for the ``faint'' relation
and 
$\Delta{J} = 3.1{\pm}0.4$~mag, $\Delta{H} = 3.8{\pm}0.5$~mag, 
$\Delta{K} = 4.3{\pm}0.6$~mag for the ``bright'' relation,
as calculated directly from the flux-calibrated 
spectral templates.  There is a large
difference in the relative magnitudes
between these two relations.  If resolved photometry is eventually obtained
for this system, such measurements could provide a means
of distinguishing which of the absolute magnitude relations proposed in
\citet{liu06} accurately characterize mid-type T dwarfs.

The origin of the 1.6~$\micron$ feature in the spectrum of 
{\namesh} is clearly revealed in Figures~\ref{fig_double} and~\ref{fig_double2}: 
it is a combination of FeH absorption in the
M dwarf primary and {\meth} absorption in the T dwarf secondary.
Specifically, the relatively sharp $H$-band flux peak in the spectrum 
of the T dwarf secondary blueward of the 1.6~$\micron$ {\meth}
band contributes light
to the 1.55--1.6~$\micron$ spectrum of the composite system.  This is
on the blue end of the 1.55-1.65~$\micron$ FeH absorption
band, producing a distinct ``dip'' feature.   
Similarly, the apparently brighter 1.2--1.35~$\micron$
flux in the spectrum of {\namesh} can be attributed to the
T dwarf companion, which exhibits a
narrow $J$-band peak between strong 1.1~$\micron$ and 1.4~$\micron$
{\wat} and {\meth} bands.  
Both spectral features are therefore
unique to binaries containing late-type M and L dwarf
primaries (in which FeH is prominent) and T dwarf secondaries.

\section{Discussion}

\subsection{Is {\namesh} an M dwarf plus T dwarf Binary?}

It may be concluded from the analysis above that the
near-infrared spectrum of {\namesh}, and in particular the subtle
feature observed at 1.6~$\micron$,
can be accurately reproduced by assuming that this 
source is an unresolved M8.5 + T5 binary.  But does this mean
that {\namesh} actually is a binary?
Our LGS AO imaging observations failed to detect any 
faint secondaries near {\namesh} to the limits displayed in Figure~\ref{fig_ao}. 
Based on the ``bright'' MKO
$M_K$/spectral type relation of \citet{liu06} and the $K_s$/$K$ filter transformations
of \citet{ste04}, the measured upper limits rule out a T5 
companion wider than a projected separation of $\sim$0$\farcs$33, or roughly
8.3 AU at the estimated distance of {\namesh} (see below).
This is a relatively weak constraint given that less than 25\% of known very low mass binaries
have projected separations at least this wide \citep{me06ppv}.  Furthermore, {\namesh} could have been 
observed in an unfortunate geometry, as was originally the case for the L dwarf binary Kelu~1 
\citep{mar99a,liu05,gel06}.  
On the other hand, if the physical separation of 
the {\namesh} system is significantly smaller than indicated by 
the imaging observations,
high resolution spectroscopic monitoring could potentially
reveal radial velocity signatures, although this depends critically
on the component masses of this system.  Indeed, the determination of
a spectroscopic orbit  in combination with the component spectral types
deduced here would provide both mass and age constraints for this system, making
it a potentially powerful benchmark test for evolutionary models. 

\begin{figure}
\epsscale{0.9}
\plotone{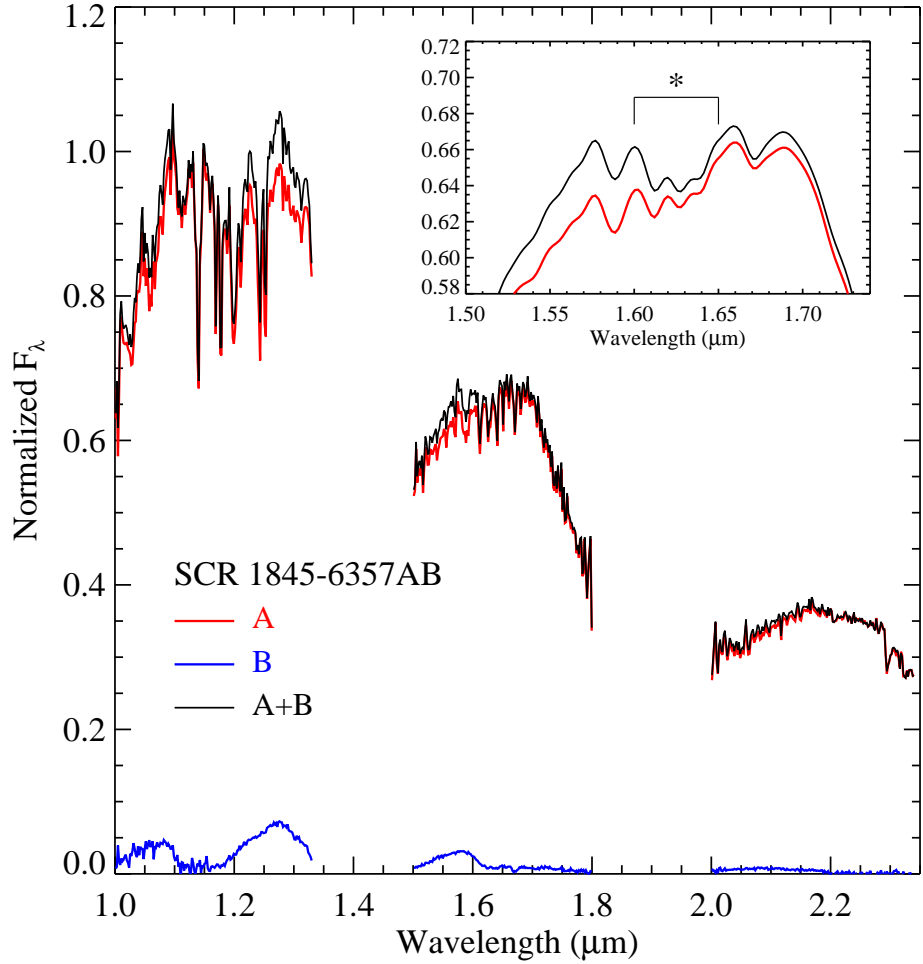}
\caption{
Component spectra of SCR 1845$-$6357AB from \citet{kas07}.
The M8.5 primary (red lines) and T6 secondary (blue lines) spectra are
scaled according to the relative $H$-band component 
photometry as reported 
by Kasper et al.  The sum of the component spectra (black lines)
shows a slight increase in both 1.25--1.35~$\micron$ and 1.55--1.6~$\micron$
flux. The inset box shows a close-up of
the primary and composite spectra in the 1.5--1.75~$\micron$ region,
where their spectral resolutions have been reduced to match that of the
SpeX prism data.  A weak 1.6~$\micron$ dip, similar to that seen in the
spectrum of {\namesh}, is also found to be present in the composite SCR~1845$-$6357
spectrum.
\label{fig_scr1845}}
\end{figure}

An alternative test of the binary hypothesis for {\namesh}
is to identify similar
spectral traits in a comparable binary system.  Fortunately, 
one such system is
known: the M8.5 + T6 binary SCR 1845$-$6357 \citep{ham04,bil06,mon06}.
This nearby (3.85$\pm$0.02~pc; \citealt{hen06}), well-resolved binary
(angular separation of 1$\farcs$1) 
has individually classified components based
on resolved spectroscopy \citep{kas07}.  More importantly, the relative
near-infrared magnitudes of this system ($\Delta{J} = 3.68{\pm}0.03$~mag, 
$\Delta{H} = 4.20{\pm}0.04$~mag, 
$\Delta{K} = 5.12{\pm}0.03$~mag; \citealt{kas07}) are somewhat larger than
but consistent with the estimated
relative magnitudes of the putative {\namesh} system.  
Figure~\ref{fig_scr1845} displays the component spectra
of this system, scaled to their relative $H$-band 
magnitudes,\footnote{The $J$-band portion of the spectrum of 
SCR~1845$-$6357A shown here
is slightly reduced relative to the $H$- and $K_s$-band spectra 
as shown to Figure~2 in \citet{kas07}. The relative 
flux calibration between spectral orders applied
in that study did not account for missing data over 1.33--1.50~$\micron$, 
slightly inflating the flux levels
in the $J$-band.  A recalibration of this spectrum was made by scaling each order by
a constant factor to match the SpeX prism spectrum of 2MASS~J1124+3808,
which has a similar $J-K_s$ color (1.14$\pm$0.03 versus 1.06$\pm$0.03 for SCR~1845$-$6357
from \citealt{kas07}) and optical spectral type (M8.5).
Such recalibration is not necessary for the SCR 1845$-$6357B spectrum due to the strong 1.35~$\micron$ {\meth} and 1.4~$\micron$ {\wat} 
bands in this source.  The recalibration of the SCR 1845$-$6357A $J$-band spectrum
does not affect the analysis
presented here, which depends solely on the relative $H$-band scaling of
the component spectra.}
as well as the sum of the component spectra.  The composite spectrum shows a relative
increase in spectral flux as compared to the primary in both the 1.2--1.35~$\micron$ 
and 1.55--1.6~$\micron$ regions.  Indeed, the latter gives rise to the same
``dip'' feature observed in the $H$-band spectrum of {\namesh},
particularly when the SCR 1845$-$6357AB data are reduced in resolution
to match that of the SpeX prism data (inset box in Figure~\ref{fig_scr1845}).
The presence of this feature in the composite spectrum of a known M dwarf plus T dwarf
binary lends some confidence to the conclusion that {\namesh}
is itself an M dwarf plus T dwarf binary.

Assuming then that {\namesh} is a system with M8.5 and T5 dwarf components,
it is possible to characterize the physical properties of these components in some detail
based on the analysis in $\S$~3.3.
Synthetic component $JHK$ magnitudes on the MKO system 
assuming the ``bright'' $M_K$/spectral type relation of \citet{liu06}
were computed from the best-fitting binary templates ($\sigma^2 < 0.1$) and 
are listed in Table~\ref{tab_component}.
The M dwarf primary is only slightly fainter than the composite source,
while the T dwarf companion is exceptionally faint, $J$ = 16.4$\pm$0.4~mag.
The low luminosity of the secondary, {\lbol} = -5.0$\pm$0.3 dex based
on its inferred spectral type \citep{gol04,meltbinary}, 
suggests that {\namesh} could 
have a relatively low system mass ratio ($q \equiv$ M$_2$/M$_1$).
However, the mass ratio depends critically on the age of the system, for
which the analysis presented above provides no robust constraints.  
Using the evolutionary models of \citet{bur97}
and component luminosities as listed in Table~\ref{tab_component},
primary and secondary mass estimates for ages of 1, 5 and 10~Gyr were derived.  
If {\namesh} is an older
system, as suggested by its large {\vtan}, its inferred mass ratio  
$q$ $>$ 0.8 is consistent with the typical mass ratios 
of very low mass binaries in the field (e.g., \citealt{pall07}). 
Based on the primary's photometry and spectral type,
and the $M_J$/spectral type relation of \citet{cru03},
a distance of 25$\pm$3~pc is estimated for the {\namesh} system.

\subsection{On the Identification of  M dwarf plus T dwarf Binaries from Composite Near-Infrared Spectra}

\begin{figure}
\epsscale{0.9}
\plotone{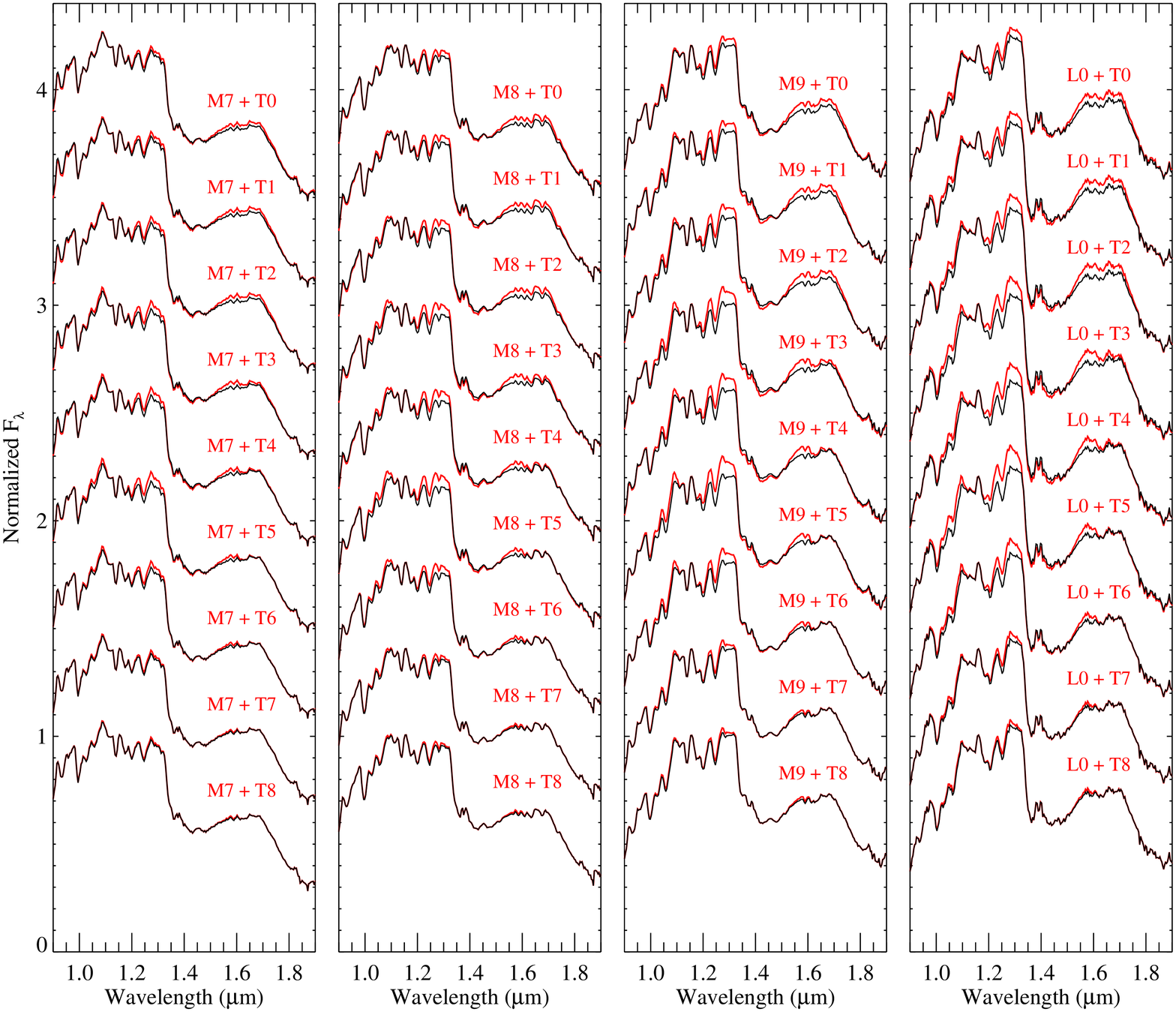}
\caption{
Simulated M7--L0 plus T dwarf binary spectra (red lines),
based on the ``bright'' $M_K$/spectral type relation of \citet{liu06}.
The primaries shown (black lines) are VB~8 (M7), VB~10 (M8),
LHS~2924 (M9) and 
2MASS~J0345+2540 (L0; see Table~\ref{tab_templates}).  The 
T dwarf secondaries are the standards defined in
\citet{meclass2}.  
Binary templates were constructed as described in $\S$~3.3,
and all spectra are normalized in the 1.12--1.17~$\micron$ region
where the T dwarf companions contribute minimal flux.
\label{fig_mtsim}}
\end{figure}

The subtlety of the peculiar features present in the composite
spectra of {\namesh} and SCR 1845$-$6357 is due entirely to the considerable
difference in flux between their M and T dwarf components.  Yet in both cases the
1.6~$\micron$ feature, indicating the presence of a T dwarf companion, can
be discerned.  But for how early of an primary can a
binary with a T dwarf companion be identified in this manner,
and what variety of 
T dwarf companions can be discerned in such systems?
To examine these questions,
Figure~\ref{fig_mtsim} displays binary spectral templates
for four primary types---M7, M8, M9 and
L0---combined with T0--T8 dwarf secondaries.  
For all cases, the 1.6~$\micron$ feature
is most pronounced when the secondary is a mid-type T dwarf, spectral types
T3--T5.  This is due to
a tradeoff in the sharpness of the $H$-band flux peak in this component
(i.e., the strength of 1.6~$\micron$ {\meth} absorption, which deepens
with later spectral types) and its
brightness relative to the primary.  
Not surprisingly, the 1.6~$\micron$ feature is more pronounced in binaries
with later-type primaries, making it a useful multiplicity diagnostic
for L dwarf + T dwarf systems (such as SDSS~J0805+4812) but far
more subtle in systems with M dwarf primaries.  Indeed, 
the spectra in Figure~\ref{fig_mtsim} suggest that this feature
is basically undetectable in binaries with M7 and earlier-type 
primaries.  {\namesh} and SCR 1845$-$6357 probably contain the earliest-type
primaries for which a T dwarf secondary could be identified solely from 
their composite near-infrared spectra.

It is also important to consider the other prominent spectral peculiarity
caused by the presence of a T dwarf companion, the slight
increase in flux at 1.3~$\micron$.  
This feature increases the contrast in the 1.4~$\micron$ {\wat} band,
and therefore serves to bias {\wat} spectral indices toward later
subtypes.  This effect explains why the near-infrared classification
of {\namesh} is so much later than its optical classification (the T dwarf
secondary contributes negligible flux in the optical). Figure~\ref{fig_mtsim} shows
that the 1.3~$\micron$ flux increase can be discerned for systems with early-
and mid-type T dwarf companions. 
While it is again more pronounced
for systems with later-type primaries, it is still present (but subtle)
in the spectra of systems with M7 primaries.  A source with unusually strong absorption
at 1.35~$\micron$, or equivalently with a near-infrared spectral
type that is significantly later than its optical spectral type, may harbor a 
T dwarf companion.  However, other physical effects, notably reduced condensate
opacity (e.g., \citealt{me1126}), can also give rise to this spectral peculiarity.  
Hence, both the contrast of the 1.4~$\micron$ {\wat} band
and the presence of the 1.6~$\micron$ dip should be considered together 
as indicators of an unresolved T dwarf companion.

Detecting the near-infrared spectral signature of a T dwarf companion need not be limited
to low-resolution observations.  While the dip feature at 1.6~$\micron$ is less
pronounced in the higher-resolution composite spectrum of SCR~1845$-$6357AB from \citet{kas07},
individual {\meth} lines may still be distinguishable amongst the many FeH and {\wat}
lines present in the same spectral region.
It may also be possible to identify {\meth} lins amongst the forest
of {\wat} lines in the 1.30-1.35~$\micron$ region (e.g. \citealt{bar06}).  Such detections
require significantly higher resolutions, of order {\ldl} $\approx$ 20,000 or more, due to the 
substantial overlap of the many molecular features present at these wavelengths (e.g., \citealt{mcl07}).
Furthermore, an improved line list for the {\meth} molecule 
may be needed \citep{sha07}.
Yet such observations have the potential to provide an additional check on the
existence and characteristics of mid-type T dwarf companions in binaries with  
late-M/L dwarf primaries.
	
Relevant to the identification of late-type
M dwarf plus T dwarf binaries  from composite
near-infrared spectra is  the number of such systems that
are expected to exist.  As a rough estimate, we examined the results of 
the Monte Carlo mass function and multiplicity simulations presented in
\citet{meltbinary}.  Using the baseline assumptions of these simulations--a mass
function that scales as $\frac{dN}{d{\rm M}} \propto {\rm M}^{-0.5}$, a component mass range of
 0.01 $\leq$ M $\leq$ 0.1~M$_{\sun}$,
a flat age distribution over 10~Gyr, the \citet{bar03} evolutionary models,
and a binary mass ratio distribution that scales as
$f(q) \propto q^{1.8}$ (see \citealt{pall07})---we found that 12-14\% of 
binaries with M8--L0 primaries are predicted to contain a T3--T5 secondary;
i.e., detectable with composite near-infrared spectroscopy.  These are primarily older systems
whose components that just straddle the hydrogen burning minimum mass limit 
($\sim$0.07~M$_{\sun}$; \citealt{cha00a}).  The overall binary fraction 
of very low mass stars and brown dwarfs has been variously estimated to lie in
the 10--35\% range (e.g. \citealt{bou03,clo03,bas06,mehst2,meltbinary,pall07,kra08}), 
and is thus currently uncertain by over a factor of three.  However, within this range the Monte Carlo simulations
predict that 1-5\% of {\em all} M8--L0 dwarfs harbor a T3--T5 dwarf companion.  
While this percentage
is small, in a given magnitude-limited survey there may be a similar number
of T dwarf companions in these relatively bright systems as compared to faint, isolated T dwarfs.
Such companions, based on the analysis above, can be reasonably well-characterized without the need of resolved imaging.

There are many other variables that must be considered if the binary spectral template technique
described here and in \citet{me0805} is to be used to determine
accurate binary statistics for very low mass stars and brown dwarfs. 
Component peculiarities, such as unusual surface gravities or cloud variations;  
intrinsic scatter in absolute magnitude/spectral
type relations; magnetic- or weather-induced photometric variability;
the detailed properties of the still poorly-constrained L dwarf/T dwarf transition;
and the possible presence of tertiary components all contribute 
in constraining the variety of systems that can be identified from composite near-infrared
spectroscopy.  Furthermore, because brown dwarfs cool over their lifetimes, 
the detectability of binaries based on component spectral types does not map uniquely to the
detectability of 
binaries based on their mass ratios and ages, resulting in complex
selection biases.  These issues will be addressed in a future
publication.

\section{Conclusions}

We have found that subtle peculiarities observed
in the near-infrared spectrum of {\namesh}, in particular a characteristic
bowl-shaped
dip at 1.6~$\micron$, indicate the presence of a
mid-type T dwarf companion.  This companion is 
unresolved in LGS AO imaging observations (including the first application
of aperture mask interferometry with LGS AO), indicating a maximum projected
separation of 8.3~AU at the time of observations.  
The binary scenario not only provides a
simple and straightforward explanation
for the 1.6~$\micron$ feature---also present in the composite
spectrum of the known M8.5 + T6 binary SCR 1845$-$6357---but also resolves
the discrepancy between the optical and near-infrared classifications
of {\namesh}.   Furthermore, empirical binary templates composed
of ``normal'' M dwarf plus T dwarf pairs provide
a far superior match to the overall near-infrared
spectral energy distribution
of {\namesh} than any single comparison source.
The hypothesis that {\namesh} is an unresolved binary is therefore
compelling, and could potentially be verified
through radial velocity monitoring observations.  
In addition, we estimate that roughly 1-5\% of all late-type
M dwarfs may harbor a mid-type T dwarf companion that could similarly be identified
and characterized using low resolution near-infrared spectroscopy and binary spectral template
analysis.

\acknowledgements

The authors acknowledge telescope operator Paul Sears
and instrument specialist John Rayner at IRTF, and 
Al Conrad, Randy Campbell, Jason McIlroy, and Gary Punawai at Keck,
for their assistance during the
observations.  We also thank Markus Kasper for providing
the spectral data for SCR 1845$-$6357 and  
Sandy Leggett, Dagny Looper and Kevin Luhman
for providing a portion of the SpeX prism spectra used in the 
binary spectral template analysis.
Our anonymous referee provided a helpful and very prompt 
critique of the original manuscript. 
MCL and TJD acknowledge support for this work from NSF
grant AST-0507833 and an Alfred P. Sloan Research Fellowship.
This publication makes
use of data from the Two Micron All Sky Survey, which is a joint
project of the University of Massachusetts and the Infrared
Processing and Analysis Center, and funded by the National
Aeronautics and Space Administration and the National Science
Foundation. 2MASS data were obtained from the NASA/IPAC Infrared
Science Archive, which is operated by the Jet Propulsion
Laboratory, California Institute of Technology, under contract
with the National Aeronautics and Space Administration.
This research has benefitted from the M, L, and T dwarf compendium housed at DwarfArchives.org and maintained by Chris Gelino, Davy Kirkpatrick, and Adam Burgasser;
the VLM Binaries Archive maintained by Nick Siegler at
\url{http://www.vlmbinaries.org}; and
the SpeX Prism Spectral Libraries, maintained by Adam Burgasser at 
\url{http://www.browndwarfs.org/spexprism}.
The authors wish to recognize and acknowledge the 
very significant cultural role and reverence that 
the summit of Mauna Kea has always had within the 
indigenous Hawaiian community.  We are most fortunate 
to have the opportunity to conduct observations from this mountain.

Facilities: \facility{IRTF~(SpeX)}, \facility{Keck~(NIRC2,LGS)}

\clearpage

\begin{deluxetable}{llllcl}
\tabletypesize{\footnotesize}
\tablecaption{SpeX Spectral Templates. \label{tab_templates}}
\tablewidth{0pt}
\tablehead{
 & & \multicolumn{2}{c}{Spectral Types} \\
\cline{3-4}
\colhead{Name} &
\colhead{2MASS Designation\tablenotemark{a}} &
\colhead{Optical} &
\colhead{NIR} &
\colhead{2MASS $J$} &
\colhead{References\tablenotemark{b}}  \\
}
\startdata
SDSS J0000+2554 & J00001354+2554180 & \nodata & T4.5 & 15.06$\pm$0.04 & {\bf 1};2 \\
2MASS J0034+0523 & J00345157+0523050 & \nodata & T6.5 & 15.54$\pm$0.05 & {\bf 3};1 \\
2MASS J0036+1821 & J00361617+1821104 & L3.5 & L4$\pm$1 & 12.47$\pm$0.03 & {\bf 4};2,5,6 \\
HD 3651B & J0039191+211516 & \nodata & T7.5 & 16.16$\pm$0.03 & {\bf 7};8,9,10 \\
2MASS J0050$-$3322 & J00501994$-$3322402 & \nodata & T7 & 15.93$\pm$0.07 & {\bf 11};1,12 \\
2MASS J0103+1935 & J01033203+1935361 & L6 & \nodata & 16.29$\pm$0.08 & {\bf 13};6 \\
2MASS J0117$-$3403 & J01174748$-$3403258 & L2: & \nodata & 15.18$\pm$0.04 & {\bf 56};14 \\
SDSS J0119+2403 & J01191207+2403317 & \nodata & T2 & 17.02$\pm$0.18 & {\bf 15} \\
IPMS 0136+0933 & J01365662+0933473 & \nodata & T2.5 & 13.46$\pm$0.03 & {\bf 4};16 \\
2MASS J0144$-$0716 & J01443536$-$0716142 & L5 & \nodata & 14.19$\pm$0.03 & {\bf 4};17 \\
SDSS J0151+1244 & J01514155+1244300 & \nodata & T1 & 16.57$\pm$0.13 & {\bf 3};1,18 \\
2MASS J0205+1251 & J02050344+1251422 & L5 & \nodata & 15.68$\pm$0.06 & {\bf 19};6 \\
SDSS J0207+0000 & J02074284+0000564 & \nodata & T4.5 & 16.80$\pm$0.16 & {\bf 1};18 \\
2MASS J0208+2542 & J02081833+2542533 & L1 & \nodata & 13.99$\pm$0.03 & {\bf 4};6 \\
SIPS J0227$-$1624  & J02271036$-$1624479 & L1 & \nodata & 13.57$\pm$0.02 & {\bf 4};20 \\
2MASS J0228+2537 & J02281101+2537380 & L0: & L0 & 13.84$\pm$0.03 & {\bf 4};14,21 \\
GJ 1048B & J02355993$-$2331205 & L1 & L1 & \nodata & {\bf 4};22 \\
2MASS J0241$-$1241 & J02415367$-$1241069 & L2: & \nodata & 15.61$\pm$0.07 & {\bf 56};14 \\
2MASS J0243$-$2453 & J02431371$-$2453298 & \nodata & T6 & 15.38$\pm$0.05 & {\bf 3};1,23 \\
SDSS J0247$-$1631 & J02474978$-$1631132 & \nodata & T2$\pm$1.5 & 17.19$\pm$0.18 & {\bf 15} \\
SO 0253+1625 & J02530084+1652532 & M7 & \nodata & 8.39$\pm$0.03 & {\bf 4};24,25 \\
DENIS J0255$-$4700 & J02550357$-$4700509 & L8 & L9 & 13.25$\pm$0.03 & {\bf 1};26,27 \\
2MASS J0310+1648 & J03105986+1648155 & L8 & L9 & 16.03$\pm$0.08 & {\bf 28};1,6 \\
SDSS J0325+0425 & J03255322+0425406 & \nodata & T5.5 & 16.25$\pm$0.14 & {\bf 15} \\
2MASS J0328+2302 & J03284265+2302051 & L8 & L9.5 & 16.69$\pm$0.14 & {\bf 4};2,6 \\
LP 944$-$20 & J03393521$-$3525440 & M9 & \nodata & 10.73$\pm$0.02 & {\bf 4} \\
2MASS J0345+2540 & J03454316+2540233 & L0 & L1$\pm$1 & 14.00$\pm$0.03 & {\bf 29};2,30,31 \\
SDSS J0351+4810 & J03510423+4810477 & \nodata & T1$\pm$1.5 & 16.47$\pm$0.13 & {\bf 15} \\
2MASS J0407+1514 & J04070885+1514565 & \nodata & T5 & 16.06$\pm$0.09 & {\bf 3};1 \\
2MASS J0415$-$0935 & J04151954$-$0935066 & T8 & T8 & 15.70$\pm$0.06 & {\bf 3};1,23,32 \\
2MASS J0439$-$2353 & J04390101$-$2353083 & L6.5 & \nodata & 14.41$\pm$0.03 & {\bf 28};14 \\
2MASS J0510$-$4208 & J05103520$-$4208140 & \nodata & T5 & 16.22$\pm$0.09 & {\bf 33} \\
2MASS J0516$-$0445 & J05160945$-$0445499 & \nodata & T5.5 & 15.98$\pm$0.08 & {\bf 4};1,34 \\
2MASS J0559$-$1404 & J05591914$-$1404488 & T5 & T4.5 & 13.80$\pm$0.02 & {\bf 1};32,35 \\
2MASS J0602+4043 & J06020638+4043588 & \nodata & T4.5 & 15.54$\pm$0.07 & {\bf 33} \\
LEHPM 2$-$461 & J06590991$-$4746532	& M6.5 & M7 & 13.64$\pm$0.03 & {\bf 4};36,37 \\
2MASS J0727+1710 & J07271824+1710012 & T8 & T7 & 15.60$\pm$0.06 & {\bf 11};23,32 \\
2MASS J0729$-$3954 & J07290002$-$3954043 & \nodata & T8 & 15.92$\pm$0.08 & {\bf 33} \\
2MASS J0755+2212 & J07554795+2212169 & T6 & T5 & 15.73$\pm$0.06 & {\bf 1};23,32 \\
SDSS J0758+3247 & J07584037+3247245 & \nodata & T2 & 14.95$\pm$0.04 & {\bf 4};1,2 \\
SSSPM 0829$-$1309 & J08283419$-$1309198 & L2 & \nodata & 12.80$\pm$0.03 & {\bf 38};39,40 \\
SDSS J0830+4828 & J08300825+4828482 & L8 & L9$\pm$1 & 15.44$\pm$0.05 & {\bf 4};18,27 \\
SDSS J0837$-$0000 & J08371718$-$0000179 & T0$\pm$2 & T1 & 17.10$\pm$0.21 & {\bf 33};1,32,41 \\
2MASS J0847$-$1532 & J08472872$-$1532372 & L2 & \nodata & 13.51$\pm$0.03 & {\bf 42};14 \\
SDSS J0858+3256 & J08583467+3256275 & \nodata & T1 & 16.45$\pm$0.12 & {\bf 15} \\
SDSS J0909+6525 & J09090085+6525275 & \nodata & T1.5 & 16.03$\pm$0.09 & {\bf 15} \\
2MASS J0939$-$2448 & J09393548$-$2448279 & \nodata & T8 & 15.98$\pm$0.11 & {\bf 1};12 \\
2MASS J0949$-$1545 & J09490860$-$1545485 & \nodata & T2 & 16.15$\pm$0.12 & {\bf 1};12 \\
2MASS J1007$-$4555 & J10073369$-$4555147 & \nodata & T5 & 15.65$\pm$0.07 & {\bf 33} \\
2MASS J1010$-$0406 & J10101480$-$0406499 & L6 & \nodata & 15.51$\pm$0.06 & {\bf 19} \\
HD 89744B & J10221489+4114266 & L0 & L (early) & 14.90$\pm$0.04 & {\bf 4};43 \\
SDSS J1039+3256 & J10393137+3256263 & \nodata & T1 & 16.41$\pm$0.15 & {\bf 15} \\
2MASS J1047+2124 & J10475385+2124234 & T7 & T6.5 & 15.82$\pm$0.06 & {\bf 4};1,32,44 \\
SDSS J1048+0111 & J10484281+0111580 & L1 & L4 & 12.92$\pm$0.02 & {\bf 4};45,46 \\
SDSS J1052+4422 & J10521350+4422559 & \nodata & T0.5$\pm$1 & 15.96$\pm$0.10 & {\bf 4};15 \\
Wolf 359 & J10562886+0700527 & M6 & \nodata & 7.09$\pm$0.02 & {\bf 4} \\
2MASS J1104+1959 & J11040127+1959217 & L4 & \nodata & 14.38$\pm$0.03 & {\bf 3};14 \\
2MASS J1106+2754 & J11061197+2754225 & \nodata & T2.5 & 14.82$\pm$0.04 & {\bf 33} \\
SDSS J1110+0116 & J11101001+0116130 & \nodata & T5.5 & 16.34$\pm$0.12 & {\bf 11};1,18 \\
2MASS J1114$-$2618 & J11145133$-$2618235 & \nodata & T7.5 & 15.86$\pm$0.08 & {\bf 11};1,12 \\
2MASS J1122$-$3512 & J11220826$-$3512363 & \nodata & T2 & 15.02$\pm$0.04 & {\bf 1};12 \\
2MASS J1124+3808 & J11240487+3808054 & M8.5 & \nodata & 12.71$\pm$0.02 & {\bf 3};14 \\
SDSS J1206+2813 & J12060248+2813293 & \nodata & T3 & 16.54$\pm$0.11 & {\bf 15} \\
SDSS J1207+0244 & J12074717+0244249 & L8 & T0 & 15.58$\pm$0.07 & {\bf 33};1,45 \\
2MASS J1209$-$1004 & J12095613$-$1004008 & \nodata & T3 & 15.91$\pm$0.07 & {\bf 3};1,27 \\
SDSS J1214+6316 & J12144089+6316434 & \nodata & T3.5$\pm$1 & 16.59$\pm$0.12 & {\bf 15} \\
2MASS J1217$-$0311 & J12171110$-$0311131 & T7 & T7.5 & 15.86$\pm$0.06 & {\bf 11};1,32,44 \\
2MASS J1221+0257 & J12212770+0257198 & L0 & \nodata & 13.17$\pm$0.02 & {\bf 4};47 \\
2MASS J1231+0847 & J12314753+0847331 & \nodata & T5.5 & 15.57$\pm$0.07 & {\bf 3};1 \\
2MASS J1237+6526 & J12373919+6526148 & T7 & T6.5 & 16.05$\pm$0.09 & {\bf 48};1,32,44 \\
SDSS J1254$-$0122 & J12545393$-$0122474 & T2 & T2 & 14.89$\pm$0.04 & {\bf 3};1,32,44 \\
2MASS J1324+6358 & J13243559+6358284 & \nodata & T2 & 15.60$\pm$0.07 & {\bf 33} \\
SDSS J1346$-$0031 & J13464634$-$0031501 & T7 & T6.5 & 16.00$\pm$0.10 & {\bf 11};1,32,49 \\
SDSS J1358+3747 & J13585269+3747137 & \nodata & T4.5$\pm$1 & 16.46$\pm$0.09 & {\bf 15} \\
2MASS J1404$-$3159 & J14044941$-$3159329 & \nodata & T2.5 & 15.60$\pm$0.06 & {\bf 33} \\
LHS 2924 & J14284323+3310391 & M9 & \nodata & 11.99$\pm$0.02 & {\bf 29} \\
SDSS J1435+1129 & J14355323+1129485 & \nodata & T2$\pm$1 & 17.14$\pm$0.23 & {\bf 15} \\
2MASS J1439+1929 & J14392836+1929149 & L1 & \nodata & 12.76$\pm$0.02 & {\bf 3};31 \\
SDSS J1439+3042 & J14394595+3042212 & \nodata & T2.5 & 17.22$\pm$0.23 & {\bf 15} \\
Gliese 570D & J14571496$-$2121477 & T7 & T7.5 & 15.32$\pm$0.05 & {\bf 3};1,32,50 \\
2MASS J1503+2525 & J15031961+2525196 & T6 & T5 & 13.94$\pm$0.02 & {\bf 3};1,32,51 \\
2MASS J1506+1321 & J15065441+1321060 & L3 & \nodata & 13.37$\pm$0.02 & {\bf 28};52 \\
2MASS J1507$-$1627 & J15074769$-$1627386 & L5 & L5.5 & 12.83$\pm$0.03 & {\bf 28};2,5,6 \\
SDSS J1511+0607 & J15111466+0607431 & \nodata & T0$\pm$2 & 16.02$\pm$0.08 & {\bf 15} \\
2MASS J1526+2043 & J15261405+2043414 & L7 & \nodata & 15.59$\pm$0.06 & {\bf 3};6 \\
2MASS J1546$-$3325 & J15462718$-$3325111 & \nodata & T5.5 & 15.63$\pm$0.05 & {\bf 4};1,23 \\
2MASS J1615+1340 & J16150413+1340079 & \nodata & T6 & 16.35$\pm$0.09 & {\bf 33} \\
SDSS J1624+0029 & J16241436+0029158 & \nodata & T6 & 15.49$\pm$0.05 & {\bf 11};1,53 \\
2MASS J1632+1904 & J16322911+1904407 & L8 & L8 & 15.87$\pm$0.07 & {\bf 28};1,31 \\
2MASS J1645$-$1319 & J16452211$-$1319516 & L1.5 & \nodata & 12.45$\pm$0.03 & {\bf 4};54 \\
VB 8 & J16553529$-$0823401 & M7 & \nodata & 9.78$\pm$0.03 & {\bf 4} \\
SDSS J1750+4222 & J17502385+4222373 & \nodata & T2 & 16.47$\pm$0.10 & {\bf 1};2 \\
SDSS J1750+1759 & J17503293+1759042 & \nodata & T3.5 & 16.34$\pm$0.10 & {\bf 3};1,18 \\
2MASS J1754+1649 & J17545447+1649196 & \nodata & T5 & 15.81$\pm$0.07 & {\bf 4} \\
SDSS J1758+4633 & J17580545+4633099 & \nodata & T6.5 & 16.15$\pm$0.09 & {\bf 11};1,2 \\
2MASS J1807+5015 & J18071593+5015316 & L1.5 & L1 & 12.93$\pm$0.02 & {\bf 4};14,21 \\
2MASS J1828$-$4849 & J18283572$-$4849046 & \nodata & T5.5 & 15.18$\pm$0.06 & {\bf 3};1 \\
2MASS J1901+4718 & J19010601+4718136 & \nodata & T5 & 15.86$\pm$0.07 & {\bf 3};1 \\
VB 10 & J19165762+0509021 & M8 & \nodata & 9.91$\pm$0.03 & {\bf 3} \\
2MASS J2002$-$0521 & J20025073$-$0521524 & L6 & \nodata & 15.32$\pm$0.05 & {\bf 4};55 \\
SDSS J2028+0052 & J20282035+0052265 & L3 & \nodata & 14.30$\pm$0.04 & {\bf 3};45 \\
LHS 3566 & J20392378$-$2926335 & M6 & \nodata & 11.36$\pm$0.03 & {\bf 3} \\
2MASS J2049$-$1944 & J20491972$-$1944324 & M7.5 & \nodata & 12.85$\pm$0.02 & {\bf 3} \\
SDSS J2052$-$1609 & J20523515$-$1609308 & \nodata & T1$\pm$1 & 16.33$\pm$0.12 & {\bf 4,15} \\
2MASS J2057$-$0252 & J20575409$-$0252302 & L1.5 & L1.5 & 13.12$\pm$0.02 & {\bf 3};14,46 \\
2MASS J2107$-$0307 & J21073169$-$0307337 & L0 & \nodata & 14.20$\pm$0.03 & {\bf 3};14 \\
SDSS J2124+0100 & J21241387+0059599 & \nodata & T5 & 16.03$\pm$0.07 & {\bf 15};1,2 \\
2MASS J2132+1341 & J21321145+1341584 & L6 & \nodata & 15.80$\pm$0.06 & {\bf 59};55 \\
2MASS J2139+0220 & J21392676+0220226 & \nodata & T1.5 & 15.26$\pm$0.05 & {\bf 1};56 \\
HN Peg B & J21442847+1446077 & \nodata & T2.5 & 15.86$\pm$0.03 & {\bf 10} \\
2MASS J2151$-$2441 & J21512543$-$2441000 & L3 & \nodata & 15.75$\pm$0.08 & {\bf 56};55,57 \\
2MASS J2151$-$4853 & J21513839$-$4853542 & \nodata & T4 & 15.73$\pm$0.07 & {\bf 4};1,58 \\
2MASS J2154+5942 & J21543318+5942187 & \nodata & T6 & 15.66$\pm$0.07 & {\bf 33} \\
2MASS J2212+1641 & J22120345+1641093 & M5 & \nodata & 11.43$\pm$0.03 & {\bf 3} \\
2MASS J2228$-$4310 & J22282889$-$4310262 & \nodata & T6 & 15.66$\pm$0.07 & {\bf 3};1,34 \\
2MASS J2234+2359 & J22341394+2359559 & M9.5 & \nodata & 13.15$\pm$0.02 & {\bf 3} \\
SDSS J2249+0044 & J22495345+0044046 & L3 & L5$\pm$1.5 & 16.59$\pm$0.13 & {\bf 4};2,18,45 \\
2MASS J2254+3123 & J22541892+3123498 & \nodata & T4 & 15.26$\pm$0.05 & {\bf 3};1,23 \\
2MASS J2331$-$4718 & J23312378$-$4718274 & \nodata & T5 & 15.66$\pm$0.07 & {\bf 3};1 \\
2MASS J2339+1352 & J23391025+1352284 & \nodata & T5 & 16.24$\pm$0.11 & {\bf 1};23 \\
LEHPM 1$-$6333 & J23515012$-$2537386 & M8 & M8 & 12.47$\pm$0.03 & {\bf 4};36,40,55 \\
LEHPM 1$-$6443 & J23540928$-$3316266 & M8.5 & M8 & 13.05$\pm$0.02 & {\bf 4};36,40 \\
2MASS J2356$-$1553 & J23565477$-$1553111 & \nodata & T5.5 & 15.82$\pm$0.06 & {\bf 1};23 \\
 \enddata
\tablenotetext{a}{2MASS designations provide the sexigesimal Right Ascension and declination of each source at J2000 equinox: Jhhmmss[.]ss$\pm$ddmmss[.]s.}
\tablenotetext{b}{Reference for spectral data in boldface type, followed by citations for
source discovery and spectral classification, as listed at DwarfArchives.org.}
\tablerefs{(1) \citet{meclass2}; (2) \citet{kna04}; (3) \citet{mewide3};
(4) Burgasser et al., in preparation.; (5) \citet{rei00}; (6) \citet{kir00};
(7) \citet{mehd3651}; (8) \citet{mug06}; (9) \citet{liu07}; (10) \citet{luh07}; 
(11) \citet{metgrav}; (12) \citet{tin05}; (13) \citet{cru04}; (14) \citet{cru03};
(15) \citet{chi06}; (16) \citet{art06}; (17) \citet{lie03}; (18) \citet{geb02};
(19) \citet{rei2252}; (20) \citet{dea05}; (21) \citet{wil03}; (22) \citet{giz01};
(23) \citet{me02}; (24) \citet{tee03}; (25) \citet{hen06}; (26) \citet{mar99b};
(27) Kirkpatrick et al., in preparation; (28) \citet{meltbinary}; (29) \citet{me2200};
(30) \citet{kir97}; (31) \citet{kir99}; (32) \citet{me03opt}; (33) \citet{loo07};
(34) \citet{mewide2}; (35) \citet{me0559}; (36) \citet{por04}; (37) \citet{rui95};
(38) \citet{me1520}; (39) \citet{sch02}; (40) \citet{lod05}; (41) \citet{leg00};
(42) \citet{mce06}; (43) \citet{wil01}; (44) \citet{me99}; (45) \citet{haw02};
(46) \citet{ken04}; (47) Reid et al., in preparation; (48) \citet{lie07};
(49) \citet{tsv00}; (50) \citet{megl570d}; (51) \citet{mewide1}; (52) \citet{giz00};
(53) \citet{str99}; (54) \citet{giz02}; (55) \citet{cru07}; (56) Cruz et al., in preparation; 
(57) \citet{lie06}; (58) \citet{ell05}; (59) \citet{sie05}}
\end{deluxetable}

\begin{deluxetable}{lccc}
\tabletypesize{\footnotesize}
\tablecaption{Predicted Component Parameters for {\namesh}. \label{tab_component}}
\tablewidth{0pt}
\tablehead{
\colhead{Parameter} &
\colhead{{\namesh}A} &
\colhead{{\namesh}B} &
\colhead{Difference}  \\
}
\startdata
Spectral Type & M8.5$\pm$0.3 & T5$\pm$0.9 & \nodata \\
${J}$\tablenotemark{a} (mag) & 13.25$\pm$0.03  &  16.4$\pm$0.4 & 3.1$\pm$0.4 \\
${H}$\tablenotemark{a} (mag) & 12.61$\pm$0.03  &  16.4$\pm$0.5 & 3.8$\pm$0.5 \\
${K}$\tablenotemark{a} (mag) & 12.13$\pm$0.03  &  16.5$\pm$0.6 & 4.3$\pm$0.6 \\
{\lbol}\tablenotemark{b} & -3.48$\pm$0.10 & -5.0$\pm$0.3 & 1.5$\pm$0.3 \\
$\mu$ ($\arcsec$ yr$^{-1}$) & 0.562$\pm$0.005 & \nodata & \nodata \\
$\phi$ ($\degr$) & 205.9$\pm$0.5 & \nodata & \nodata \\
$d$\tablenotemark{c} (pc) & 25$\pm$3 & \nodata & \nodata \\
{\vtan} ({\kms}) & 67$\pm$8 & \nodata & \nodata \\
$\rho$ (AU) & $<$8.3 ($<$0$\farcs$33) & \nodata & \nodata \\
Mass (M$_{\sun}$) at 1 Gyr\tablenotemark{d} & 0.081 & 0.035 &  0.44\tablenotemark{e} \\ 
Mass (M$_{\sun}$) at 5 Gyr\tablenotemark{d} & 0.086 & 0.068 &  0.79\tablenotemark{e} \\
Mass (M$_{\sun}$) at 10 Gyr\tablenotemark{d} & 0.086 & 0.074 &  0.86\tablenotemark{e} \\
\enddata
\tablenotetext{a}{Synthetic magnitudes on the MKO system, based on 2MASS $JHK_s$ photometry for the unresolved source and binary template fits using
the ``bright'' $M_K$/spectral type relation of \citet{liu06}.}
\tablenotetext{b}{Based on the $M_{bol}$/spectral type relation of \citet{meltbinary}.}
\tablenotetext{c}{Based on the inferred $J$ magnitude and spectral type of the primary,
and the $M_J$/spectral type relation of \citet{cru03}.}
\tablenotetext{d}{Based on evolutionary models from \citet{bur97} and the estimated luminosities.}
\tablenotetext{e}{Mass ratio $q \equiv$ M$_2$/M$_1$.}
\end{deluxetable}
 
 \clearpage

\end{document}